# TEAM: A Multiple Testing Algorithm on the Aggregation Tree for Flow Cytometry Analysis

John Pura[a], Xuechan Li[a], Cliburn Chan[a], Jichun Xie[a,*]

[a] *Department of Biostatistics and Bioinformatics, Duke University School of Medicine, Durham, NC 27705*

**Abstract**

In immunology studies, flow cytometry is a commonly used multivariate single-cell assay. One key goal in flow cytometry analysis is to pinpoint the immune cells responsive to certain stimuli. Statistically, this problem can be translated into comparing two protein expression probability density functions (PDFs) before and after the stimulus; the goal is to pinpoint the regions where these two PDFs differ. In this paper, we model this comparison as a multiple testing problem. First, we partition the sample space into small bins. In each bin we form a hypothesis to test the existence of differential PDFs. Second, we develop a novel multiple testing method, called TEAM (Testing on the Aggregation tree Method), to identify those bins that harbor differential PDFs while controlling the false discovery rate (FDR) under the desired level. TEAM embeds the testing procedure into an aggregation tree to test from fine- to coarse-resolution. The procedure achieves the statistical goal of pinpointing differential PDFs to the smallest possible regions. TEAM is computationally efficient, capable of analyzing large flow cytometry data sets in much shorter time compared with competing methods. We applied TEAM and competing methods on a flow cytometry data set to identify T cells responsive to the cytomeglovirus (CMV)-pp65 antigen stimulation. TEAM successfully identified the monofunctional, bifunctional, and polyfunctional T cells while the competing methods either did not finish in a reasonable time frame or provided less interpretable results. Numerical simulations and theoretical justifications demonstrate that TEAM has asymptotically valid, powerful, and robust performance. Overall, TEAM is a computationally efficient and statistically powerful algorithm that can yield meaningful biological insights in flow cytometry studies.

[*]Principal corresponding author
 *Email addresses:* `john.pura@duke.edu` (John Pura), `xuechan.li@duke.edu` (Xuechan Li), `cliburn.chan@duke.edu` (Cliburn Chan), `jichun.xie@duke.edu` (Jichun Xie)

# 1. Introduction

Flow cytometry is a multivariate single-cell assay commonly used to characterize the immune system. A key challenge in flow cytometry is the identification of T cells that are activated by specific antigens such as a particular tumor, bacterial or viral protein. Intracellular cytokine staining (ICS) is often combined with flow cytometry to analyze the antigen-specific T cell immune response. In ICS, cells are first activated with antigen, followed by staining for cell surface molecules that define the cell phenotype, such as CD4 and/or CD8. Cells can be further stained after membrane permeabilization with fluorochrome-labeled monoclonal antibodies specific to protein markers within the cell, such as the IFN-$\gamma$, IL-2 and TNF-$\alpha$ cytokines that are only expressed after T cell activation. Hence T cells with high levels of cytokine expressions are likely to be antigen-specific. The flow cytometer quantifies the amount of antibodies bound to the cell, and hence the concentration of the protein marker that the antibody is targeting. The expression repertoire of protein markers is often used to characterize cell types at two levels – the cell surface markers define the basic cell type (*e.g.*, CD4+ T cell) and maturational stage (*e.g.*, memory T cells) while the pattern of intracellular cytokine expression defines its activation class. Quantification of different activation classes of antigen-specific T cells provides useful biological information such as the likely efficacy of a vaccine (Seder et al., 2008). However, as the relative frequency of antigen-specific cells for any particular antigen is often very low (rarely above 1% and often much lower), several hundred thousand cells per sample are typically evaluated to quantify different subpopulations of antigen-specific T cells. This gives rise to the need for statistical methods that are sensitive at detecting antigen-specific cells but guard against false positives.

As an illustrative example, we present a dataset from the External Quality Assurance Program Oversight Laboratory proficiency program by the Duke Immune Profiling Core (Staats et al., 2014). Blood samples from 11 healthy individuals were collected, and each sample was split into two parts. One was used as a negative control (cohort 1); the other was stimulated with a peptide mixture from the immunodominant cytomegalovirus pp65 protein (cohort 2). Cohort 1 contains nearly 2.4 million cells and cohort 2 has nearly 2.2 million cells. Each cell has 11 protein markers, collected for discriminating among T cell basic, maturational and functional subsets. We expect that specific T cells targeting the cytomegalovirus would be activated and show elevated expression levels in the four functional protein markers: TNF-$\alpha$, IL-2, IFN-$\gamma$, and CD107 (a protein associated with cytotoxic activity). The one- and two-dimensional marker density contour plots are presented in Figure 1.

Generally speaking, in a typical flow cytometry study, $N_1$ cells under condition 1 (cohort 1) and $N_2$ cells under condition 2 (cohort 2) are evaluated; for each cell, $p$ protein marker expressions are collected. Here $N_1$ and $N_2$ could range from $10^5$ to $10^7$, and $p$ could range from fewer than 10 up to 50. The goal is to pinpoint the regions where the multi-dimensional protein marker distributions of cohort 2 is higher than cohort 1. The characterization of cells activated under condition 2 has many applications - for example, in evaluating the immunogenicity of a vaccine. As mentioned above, the activated cells are typically present in very low relative frequencies, and we need to identify regions of the parameter space that are highly enriched for such cells.

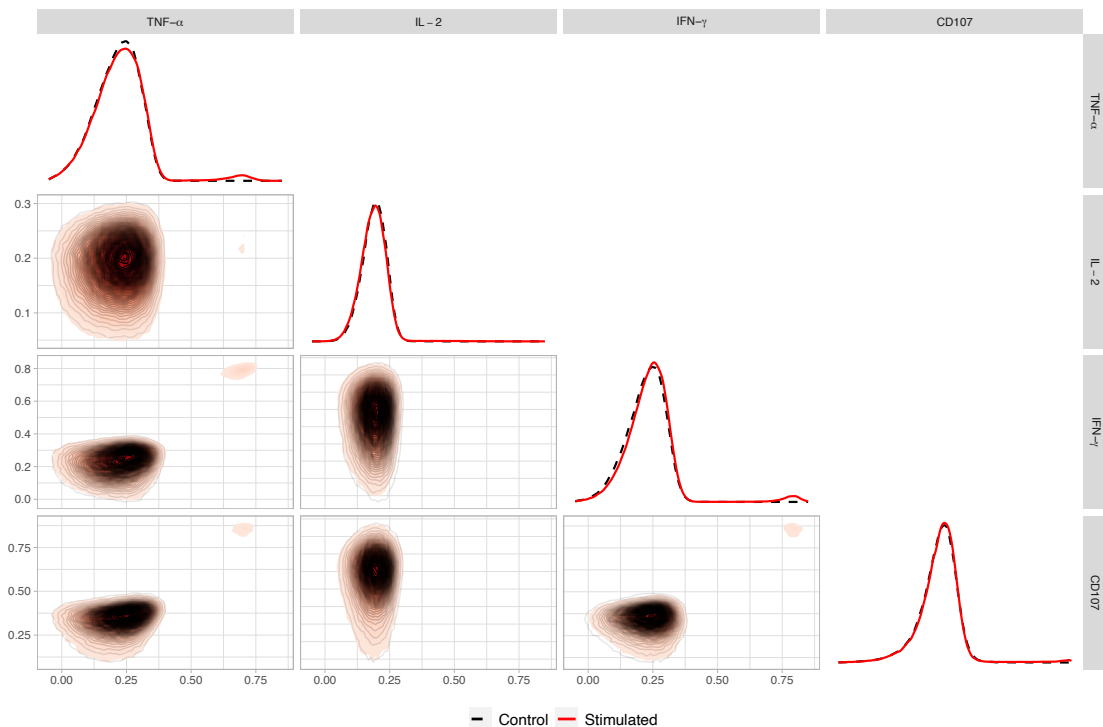

Figure 1: Univariate and bivariate distribution of activation markers on cells evaluated by FCM. Potential biological differences between cohort 1 (control; black) and cohort 2 (stimulated; red) samples can be identified through examining bivariate densities of the four functional markers: TNF-$\alpha$, IL-2, IFN-$\gamma$, and CD107. Univariate densities are plotted along the diagonals, while bivariate densities are plotted in the lower triangular region. Cohort 1 functional marker densities are displayed as grey-scale contour lines, while cohort 2 marker densities are displayed as filled, red contours.

Existing methods for this problem compare two protein expression PDFs based on the observed cells and are aimed at finding those regions which harbor differential densities. A common strategy across these methods is the initial partitioning of the sample space into discrete testing units. Roederer and Hardy (2001) proposes the frequency gating method (Roederer et al., 2001a) to compare the frequencies between stimulated and reference cells and identify regions with significant differential frequencies. The reference sample space is divided into bins containing roughly equal number of cells. The resulting partition is then applied to the stimulated sample and the ratio of the stimulated and reference cells in each bin is calculated to derive a normalized chi-square statistic for each bin. Then they use a user-defined threshold to reject those bins with large chi-square statistics. Duong (2013) uses kernel density estimation and chi-square test statistics to identify local distributional difference at any given location in multi-dimensional space. Antoniadis et al. (2015) focuses on identifying one-dimensional differential density regions. They divide the samples into equal length bins, and then use the Poisson distribution to model the number of stimulated and reference cells in each bin. Then they use a variance stabilization transformation to normalize the counts and applied existing multiple-testing procedures to identify differential density regions. More recently, Soriano and Ma (2017) propose a multi-resolution scanning (MRS) method for identifying differential density regions. The method can be viewed as a testing method embedded in the partition tree. MRS sequentially partitions the common support of two distributions into finer and finer bins. On every resolution level, each bin is coupled with a null hypothesis. MRS forms a hypothesis for each large or small bin, and asymptotically controls the false discovery rate (FDR) among all these hypotheses.

In this paper, we model the challenge of pinpointing differential regions as a multiple-testing problem, and propose a new FDR controlling procedure, called TEAM. First, TEAM partitions the multivariate sample space into bins with the finest resolution. It can accommodate different partitioning schemes. Second, TEAM embeds testing on an aggregation tree. On layer 1, within each bin, TEAM tests if the PDF of cohort 2 is higher than that of cohort 1. On higher layers, TEAM will gradually aggregate the accepted bins and test if the aggregated bins harbor differential PDFs. This fine-resolution to coarse-resolution testing structure not only boosts the testing power, but also pinpoints the regions with differential PDFs at the finest possible resolution.

Although both TEAM and MRS are both testing methods embedded in a generated tree, there are several fundamental differences between these two methods. First, TEAM is embedded in an aggregation tree. It starts bottom-up from the finest bins for testing and gradually aggregates these bins to borrow information across local regions. In contrast, MRS is embedded in the partition tree. It starts top-down from the coarsest bins for testing and gradually partitions the bins to identify local differences. Second, TEAM prioritizes identification of small bins and pinpoints the regions harboring differential PDFs at the finest possible resolutions. In contrast, MRS focuses on global visualization. For example, consider the situation where cohort 2 density is higher than cohort 1 in a very small region. Because MRS starts from large bins, if the large bin contains some small differential density region, the signal in the small region could be masked by the noise in the large bin. As a result, MRS will accept the entire region corresponding to the large bin in early stages. Furthermore, MRS will prune some accepted regions in early stages to improve computation efficiency. If the accepted large bin is pruned, the small differential region will never be identified in later stages. Third, for FDR control, TEAM only considers the null hypotheses coupled with the finest-resolution bins, which by design are non-overlapping. These fine-resolution bins stand alone and are free to be null or alternative by themselves. In contrast, MRS defines FDR based on all hypotheses coupled with bins at any resolution level so that higher-resolution bins are nested in the lower-resolution bins. This means that the hypotheses in parent and children bins are "overlapping", and a much larger number of hypotheses are considered in MRS compared to TEAM. Finally, because the candidate hypotheses have more flexible structures for TEAM than for MRS, the testing results of TEAM are generally easier to interpret. In MRS, it is possible to claim a parent bin as alternative, and yet have all of its child bins as null; it is also possible to claim a parent bin as null and have some of its child bins as alternative. The nesting structure of the MRS candidate hypotheses induces difficulties in interpretation.

The rest of the paper is organized as follows. Section 2 introduces how to model the flow cytometry data. Section 3 describes the TEAM algorithm. Section 4 focuses on analyzing the flow cytometry data to identify cells responsive to the cytomegalovirus antigen. Section 5 discusses several numerical settings and compares the performance of TEAM and MRS. Section 6 provides a summary of this paper and a general discussion on the application of TEAM to other related single-cell assays.

## 2. Model

### 2.1. Sample space partitioning

Let $\Omega$ be the multivariate protein marker sample space of the pooled cells (cohort 1 and cohort 2). Consider a partition on $\Omega$, i.e., $\cup_{i=1}^{m}\Omega_i = \Omega$, $\mu(\Omega_i \cap \Omega_j) = 0$ if $i \neq j$. We call $\Omega_1, \ldots, \Omega_m$ the leaf bins (of leaf $1, \ldots, m$). Suppose the leaf bin $\Omega_i$ consists of $n_i$ cells. The partition can be constructed in several ways. Two examples include:

- *Adaptive partition.* We order the protein markers from the largest expression variance to the smallest. We choose the marker with the largest sample variance and do a median split of the sample space along this dimension. Within each subsample, we repeat these median splits of the sample space along the dimension with the largest variance. After $\tilde{m}$ times, we will have $m = 2^{\tilde{m}}$ bins. See Roederer et al. (2001b) and Roederer et al. (2001a) for details.

- *Sequential partition.* We order the protein markers from the largest expression variance to the smallest. We first partition the first marker dimension into $\tilde{m}$ bins by its sample quantiles at level $\{1/\tilde{m}\}$, ..., $\{(\tilde{m}-1)/\tilde{m}\}$. Within each bin on the first dimension, we partition the second dimension by its sample quantiles at level $\{1/\tilde{m}\}$, ..., $\{(\tilde{m}-1)/\tilde{m}\}$. We then sequentially partition all other dimensions until all dimensions are partitioned. Please see Figure 3 for an illustration.

### 2.2. Hypotheses

Let $\tilde{X}_i$ and $X_i$ be the number of cells from cohort 1 and cohort 2 that falls into leaf bin $i$, respectively. Then $n_i = X_i + \tilde{X}_i$. We know that $\sum_{i=1}^{m} \tilde{X}_i = N_1$ and $\sum_{i=1}^{m} X_i = N_2$. Consider the problem with the fixed margins $N_1$, $N_2$, and $n_i$, $i = 1, \ldots, m$. Clearly, $\{X_1, \ldots, X_m\}$ are not mutually independent. However, for any finite $K$-dimensional vector $(X_{i1}, \ldots, X_{iK})'$, its joint distribution can be well approximated by the product of the mutually-independent binomial distributions with the $i$-th component $\text{Binom}(n_i, \theta_i)$, where

$$\theta_i = \frac{N_2 \int_{\Omega_i} f_2(y)\mathrm{d}y}{N_2 \int_{\Omega_i} f_2(y)\mathrm{d}y + N_1 \int_{\Omega_i} f_1(y)\mathrm{d}y}. \tag{1}$$

Here, $f_1$ and $f_2$ are the PDFs of cohort 1 and cohort 2, and $N_1$ and $N_2$ are the cell numbers in cohort 1 and cohort 2.

**Lemma 1.** *Let $Z_1, \ldots, Z_m$ be a sequence of independent random variables, where $Z_i$ follows $\text{Binom}(n_i, \theta_i)$. Suppose $n^{(1)} = \sup_{i=1}^{m} n_i = n_i + \delta_{n,i}$ with $\sup_{i=1}^{m}|\delta_{n,i}| = o(n^{(1)})$ and* $n^{(1)} = o(N^{1/2})$. For any constant $K$ and vector $(i_1, \ldots, i_K)$, with each element taken without replacement from $\{1, \ldots, m\}$,

$$\left| \frac{\mathsf{P}(X_{i_1} = x_{i_1}, \ldots X_{i_K} = x_{i_K} \mid N_1, N_2, n_{i_1}, \ldots, n_{i_K})}{\prod_{j=1}^{K} \mathsf{P}(Z_{i_j} = x_{i_j} \mid n_{i_j})} - 1 \right| \leq CK^2 (n^{(1)})^2 / N,$$

for some constant $C$ not depending on $K$, $n^{(1)}$, $m$, $N_1$, $N_2$, or $(x_{i1}, \ldots, x_{iK})'$.

When we justify the asymptotic properties of TEAM, only joint distributions of finite dimensional $(X_{i_1}, \ldots, X_{i_K})'$ are involved. Therefore, Lemma 1 is sufficient.

Let $\theta_0 = N_2/N$. For leaf $i$, we set up the hypothesis:

$$\mathrm{H}_{\mathrm{nul},i} : \theta_i \leq \theta_0 \quad \text{versus} \quad \mathrm{H}_{\mathrm{alt},i} : \theta_i > \theta_0. \tag{2}$$

If $\mathrm{H}_{\mathrm{alt},i}$ is true we call leaf $i$ alternative and null otherwise.

We consider one-sided tests here because they correspond to our analytical goal of locating the activated cells in cohort 2. Clearly, $\int f_s(y) dy = 1$ for both $s \in \{1, 2\}$. By the continuity of $f_1$ and $f_2$, if there exists a region where the cohort 2 density is higher, there must exist a region where the cohort 1 density is higher. For our analytical goal, we only need to find regions where the cohort 2 density is higher. Under some rare cases, researchers are interested in identifying regions with differential densities in either direction; then we can first run the one-sided test, and flip the labels of cohort 1 and cohort 2, and run the one-sided test again.

Let $\mathcal{H} = \{1, \ldots, m\}$, $\mathcal{H}_{\mathrm{nul}} = \{i : \mathrm{H}_{\mathrm{nul},i} \text{ is true}\}$, and $\mathcal{H}_{\mathrm{alt}} = \mathcal{H} \setminus \mathcal{H}_{\mathrm{nul}}$. Ideally, we would like to identify where $f_2 > f_1$, i.e., $\{y \in \Omega : f_2(y) > f_1(y)\}$. With the partitioned bins, the goal is to identify $\hat{\Omega}^+ = \cup_{i \in \mathcal{H}_{\mathrm{alt}}} \Omega_i$. To make sure that $\hat{\Omega}^+$ approximates $\Omega^+$ well, the partition should be fine enough. However, if the partition is too fine and each leaf bin only contains very few cells, the testing power will be low. Discussions on the proper theoretical choice of $m$ and $n_i$ are provided in the supplementary file.

## 3. Algorithm Description

### 3.1. Step 1: Testing on layer 1

The false discovery proportion (FDP) and false discovery rate (FDR) is defined as

$$FDP = \frac{\sum_{i \in \mathcal{H}_{\mathrm{nul}}} I(\mathrm{H}_{\mathrm{nul},i} \text{ is rejected})}{\sum_{i \in \mathcal{H}} I(\mathrm{H}_{\mathrm{nul},i} \text{ is rejected}) \vee 1}, \quad FDR = \mathsf{E}(FDP). \tag{3}$$

To control FDR, we propose the following test procedure on the bottom layer. Let $G_{0,i}^{(1)}$ be the complementary cumulative density function (CCDF) of $\mathrm{Binom}(n_i, \theta_0)$. Let $P_{0,i}^{(1)} = G_{0,i}^{(1)}(X_i)$, a random variable close to the P-value. Define the threshold $\hat{c}^{(1)}$ as

$$\hat{c}^{(1)} = \sup \left\{ a_N^{(1)} \leq c \leq \alpha : c \leq \frac{\max\left\{\sum_{i \in \mathcal{H}} I(P_{0,i}^{(1)} < c), 1\right\}}{m} \cdot \alpha \right\}. \tag{4}$$

Here $a_N^{(1)} = (m \log m)^{-1}$. If such $\hat{c}^{(1)}$ does not exist, set $\hat{c}^{(1)} = a_N^{(1)}$. We reject $\mathrm{H}_{\mathrm{nul},i}$ if $P_{0,i} \leq \hat{c}^{(1)}$. This procedure is very similar to the Benjamini-Hochberg (BH) procedure (Benjamini and Hochberg, 1995).

*3.2. Step 2: Aggregation and Testing on higher layers.*

Unlike other multiple testing methods, TEAM will continue testing after layer 1. It will aggregate the neighboring accepted leaves to parent nodes and test their parent hypotheses. The underlying assumption of TEAM is that the neighboring leaves of an alternative leaf are also likely to be alternative. This assumption is reasonable under many circumstances, especially when the PDFs $f_1$ and $f_2$ are smooth (See Proposition 1). Thus, TEAM hierarchically aggregates the neighboring leaves so that their signals can be aggregated and amplified. It is possible that an alternative region spanning multiple leaf bins is missed on low layers but will be identified on higher layers.

Figure 2 provides a toy example to illustrate how TEAM works.

- Leaf 1 and leaf 2 are accepted on layer 1, so they are aggregated as a parent node $S_1^{(2)} = \{1, 2\}$. The coupled node hypothesis is

$$\mathrm{H}_{\mathrm{nul},1}^{(2)} : \ \forall \ j \in S_1^{(2)}, \ \theta_j \leq \theta_0 \quad \text{versus} \quad \mathrm{H}_{\mathrm{alt},1}^{(2)} : \ \exists \ j \in S_1^{(2)}, \ \theta_j > \theta_0.$$

- Leaf hypothesis $\mathrm{H}_{\mathrm{nul},8}^{(1)}$ is rejected on the bottom layer. As a result, leaves 7 and 9 are aggregated on layer 2, so that $S_4^{(2)} = \{7, 9\}$. This aggregation design allows the differential peak to be captured at a lower layer and the differential shoulder to be captured at a higher layer. See the illustrated distributions in Figure 2. Aggregating the shoulder areas will likely increase power.

- Leaf 12 is left alone on the bottom layer because no more leaves are left to be aggregated. On any layer, at most 1 node will be left out.

More generally, higher layers of TEAM employ two steps: aggregation and testing.

*Aggregation.* On layer $\ell$, we aggregate the neighboring accepted child nodes on layer $\ell-1$ into the parent nodes. If the leaf bins are ordinal, the aggregation can be easily performed according to the ordinal rankings. Figure 2 illustrates a one-dimensional example, and Figure 3 illustrates a two-dimensional example. The two-dimensional example can be easily extended to multiple dimensions.

*Node hypothesis.* After aggregation, new parent nodes $S_i^{(\ell)}$ are formulated. We set up the coupled node hypothesis:

$$\mathrm{H}_{\mathrm{nul},i}^{(\ell)} : \forall \ j \in S_i^{(\ell)}, \ \theta_j \leq \theta_0 \quad \text{versus} \quad \mathrm{H}_{\mathrm{alt},i}^{(\ell)} : \exists \ j \in S_i^{(\ell)}, \ \theta_j > \theta_0. \qquad (5)$$

We test these hypotheses on layer $\ell$ and map the rejections back to the bottom layer with the finest resolution.

*Testing.* On layer $\ell$, suppose there are $m^{(\ell)}$ aggregated nodes on layer $\ell$. For $\ell \geq 2$, each node $S_i^{(\ell)}$ is the union of two child nodes on $\ell - 1$, denoted by $S_{i_1}^{(\ell-1)}$ and $S_{i_2}^{(\ell-1)}$.

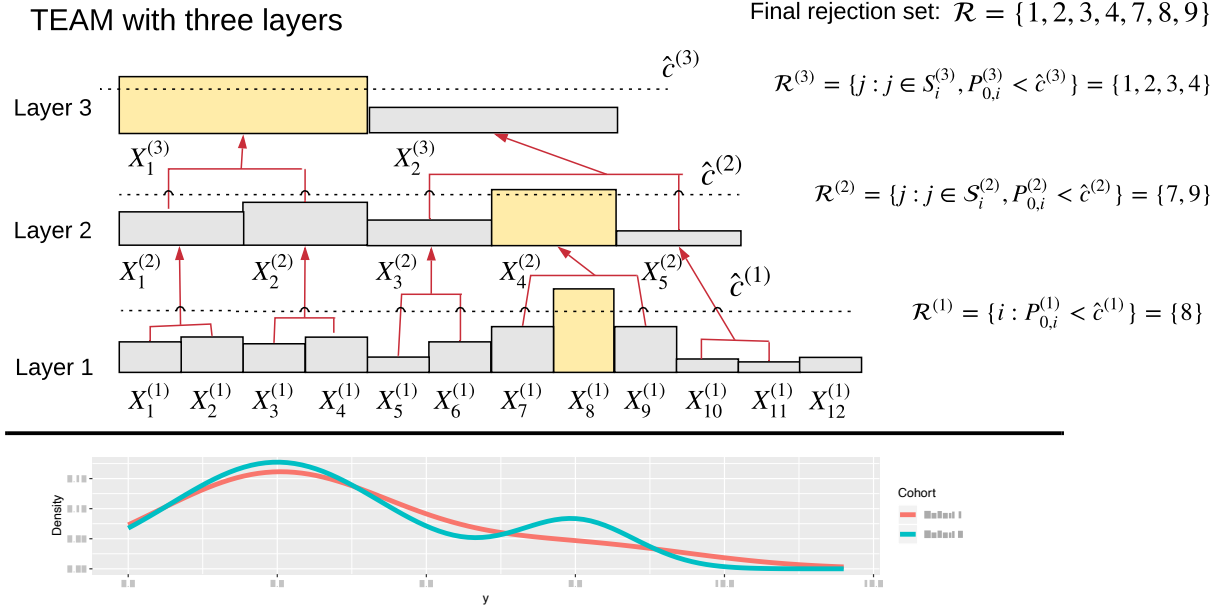

Figure 2: An illustrating example of TEAM with three layers. The non-rejected bins are aggregated at the beginning of layer 2 and layer 3, and each parent bin is coupled with a parent hypothesis. If a parent hypothesis is rejected, the rejection is mapped back to the bottom layer. For example, at the beginning of layer 2, the non-rejected leaf bin set is $\tilde{\mathcal{H}}^{(2)} = \{1, 2, 3, 4, 5, 6, 7, 9, 10, 11, 12\}$, and the parent bin set is $\mathcal{A}^{(2)} = \{\{1, 2\}, \{3, 4\}, \{5, 6\}, \{7, 9\}, \{10, 11\}\}$. On layer 2, the null hypothesis coupled with the parent bin $\{5, 6\}$ is rejected. The rejection is mapped to the bottom layer so that $H_{\text{nul},5}$ and $H_{\text{nul},6}$ are rejected.

Obviously $\mathsf{Card}(S_i^{(\ell)}) = 2^{\ell-1}$. The node bin contains $n_i^{(\ell)} = \sum_{j \in S_i^{(\ell)}} n_j^{(1)}$ samples, out of which $X_i^{(\ell)} = \sum_{j \in S_i^{(\ell)}} X_j^{(1)}$ are from cohort 2. It is easy to see that $X_i^{(\ell)} = X_{i_1}^{(\ell-1)} + X_{i_2}^{(\ell-1)}$, where $X_{i_1}^{(\ell-1)}$ and $X_{i_2}^{(\ell-1)}$ are the number of cohort 2 samples in the bins of its child node $S_{i_1}^{(\ell-1)}$ and $S_{i_2}^{(\ell-1)}$.

Similar to layer 1, we first derive a P-value-like statistic $P_{0,i}^{(\ell)} = G_{0,i}^{(\ell)}(X_i^{(\ell)}; \hat{c}^{(\ell-1)})$. Here $G_{0,i}^{(\ell)}(c; \hat{c}^{(\ell-1)})$ is defined recursively. On layer 1, $G_{0,i}(c; \hat{c}^{(0)})$ is the CCDF of $\text{Binom}(n_i^{(1)}, \theta_0)$. On layer $\ell \geq 2$, define

$$G_{0,i}^{(\ell)}(c; \hat{c}^{(\ell-1)}) = \mathsf{P}_{\theta_0}(Z_1 > c \mid G_{0,i_1}^{(\ell-1)}(Z_1; \hat{c}^{(\ell-2)}) > \hat{c}^{(\ell-1)}, G_{0,i_2}^{(\ell-1)}(Z_2; \hat{c}^{(\ell-2)}) > \hat{c}^{(\ell-1)}), \quad (6)$$

where $Z_1$ and $Z_2$ independently follows $\text{Binom}(n_{i_1}^{(\ell-1)}, \theta_0)$ and $\text{Binom}(n_{i_2}^{(\ell-1)}, \theta_0)$ respectively, and $Z = Z_1 + Z_2$.

Technically speaking, $P_{0,i}^{(\ell)}$ is not the P-value of $X_i^{(\ell)}$ because the null distribution of $X_i^{(\ell)}$ depends on the entire testing path. However, we can show that $|P_{0,i}^{(\ell)} - \breve{P}_{0,i}|/\breve{P}_{0,i}$ converges to zero in probability, where $\breve{P}_{0,i}$ is the P-value of $X_i^{(\ell)}$ (Lemma 9).

We will reject the node hypothesis $\mathrm{H}_{\mathrm{nul},i}^{(\ell)}$ if $\mathsf{P}_{0,i}^{(\ell)} \leq \hat{c}^{(\ell)}(\alpha)$, where

$$\hat{c}^{(\ell)}(\alpha) = \sup \left\{ a_N^{(\ell)} \leq c \leq \alpha : c \leq \frac{\max\left[\sum_{1 \leq i \leq m^{(\ell)}} I\{P_{0,i}^{(\ell)} \leq c\}, 1\right]}{m^{(\ell)}} \cdot \alpha \right\}, \quad (7)$$

with $a_N^{(\ell)} = (m^{(\ell)} \log m^{(\ell)})^{-1}$. If such $\hat{c}^{(\ell)}$ does not exist, set $\hat{c}^{(\ell)} = a_N^{(\ell)}$.

To map the node-level rejections to the leaves (on layer 1), we adopt an aggressive approach:

If $\mathrm{H}_{\mathrm{nul},i}^{(\ell)}$ is rejected, then $\forall\, j \in S_i^{(\ell)}, \mathrm{H}_{\mathrm{nul},j}$ are rejected.

For example, in Figure 2, we reject $\mathrm{H}_{\mathrm{nul},1}^{(3)}$. Its corresponding node is $S_1^{(3)} = \{1,2,3,4\}$. Then we reject $\mathrm{H}_{\mathrm{nul},1}$, $\mathrm{H}_{\mathrm{nul},2}$, $\mathrm{H}_{\mathrm{nul},3}$, and $\mathrm{H}_{\mathrm{nul},4}$. This aggressive approach is based on the underlying assumption that the alternative leaves are likely to cluster. See Section S1 in the supplementary file for justification.

*3.3. Stopping rule*

To control when TEAM stops, we set up a flag: flag $= 0$ for proceeding and flag $= 1$ for stopping. At the beginning of TEAM, flag $= 0$ and switches to 1 when the stopping rule is satisfied. Here are some examples of stopping rules.

- We set up a predetermined number $L$ as the maximum layer. Then TEAM will stop after $L$ layers.

- After the testing procedure on layer $\ell$ for any $\ell \geq 2$, we calculate the number of rejections on layer $\ell$. If the number is less than a prespecified level, TEAM will stop.

- After the testing procedure on layer $\ell$ for any $\ell \geq 2$, we calculate the ratio between the rejection number on layer $\ell$ and on layer $\ell - 1$. If the ratio is below a prespecified level, TEAM will stop.

*3.4. Pseudocode for TEAM*

1. Set flag $= 0$ and $\ell = 1$.

2. On layer 1, define the leaf $S_i^{(1)} = \{i\}$ for $i = 1, \ldots, m^{(1)}$ with $m^{(1)} = m$. Let the rejection set be
$$\mathcal{R}^{(1)} = \{i : P_{0,i}^{(1)} \leq \hat{c}^{(1)}\}$$
with $\hat{c}^{(1)}$ defined in (4).

3. Check the stopping rule. If it is satisfied, set flag $= 1$ and go to Step 4; otherwise, increase $\ell$ by 1 and perform the following sub-steps on layer $\ell$ ($\ell \geq 2$).

   (a) Let $\tilde{\mathcal{H}}^{(\ell)} = \mathcal{H}^{(\ell-1)} \setminus \mathcal{R}^{(\ell-1)}$.
   
   (b) Based on the predefined aggregation rule, let the node $S_i^{(\ell)} = S_{i_1}^{(\ell-1)} \cup S_{i_2}^{(\ell-1)}$ for $i = 1, \ldots, m^{(\ell)}$. Here $\mathsf{Card}(S_i^{(\ell)}) = 2^\ell - 1$ and $m^{(\ell)} = \lfloor \mathsf{Card}(\tilde{\mathcal{H}}^{(\ell)})/2^{\ell-1} \rfloor$.

(c) Set $\mathcal{H}^{(\ell)} = \cup_{i=1}^{m^{(\ell)}} S_i^{(\ell)}$, $\mathcal{H}_{\text{nul}}^{(\ell)} = \mathcal{H}^{(\ell)} \cap \mathcal{H}_{\text{nul}}$, and $\mathcal{H}_{\text{alt}}^{(\ell)} = \mathcal{H}^{(\ell)} \cap \mathcal{H}_{\text{alt}}$.

(d) Obtain the rejection set

$$\mathcal{R}^{(\ell)} = \{j : j \in S_i^{(\ell)}, \ P_{0,i}^{(\ell)} \leq \hat{c}^{(\ell)}\}.$$

with $\hat{c}^{(\ell)}$ defined in (7).

4. Let the overall rejection set be $\mathcal{R} = \cup_{i=1}^{L} \mathcal{R}^{(\ell)}$. We reject $\text{H}_{\text{nul},i}$ for all $i \in \mathcal{R}$.

## 4. Application to Antigen-Specific, T cell Activation in EQAPOL Data

### 4.1. Preprocessing

The data set was preprocessed using manual gating in FlowJo software (v9.9.6) to remove debris, doublet, and aggregate cells. To address the between-sample variation across the 11 individuals, we applied quantile normalization on a per-channel basis (Hahne et al. (2010)) to each of the negative control (cohort 1) and stimulated (cohort 2) samples prior to pooling.

### 4.2. Analysis Using TEAM

Our analysis compared the $N_1 = 2.4$ million cohort 1 cells with the $N_2 = 2.2$ million cohort 2 cells from pooled blood samples of 11 individuals. We expect the stimulation to activate the small fraction of cells that recognize the CMV pp65 antigen. Activated T cells can be identified and partitioned into activation classes based on the expression of certain effector molecules whose concentration can be measured.

We applied TEAM to identify the regions where cohort 2 protein functional marker expression PDF is higher than cohort 1 protein functional marker expressions in six sub-analyses. In each sub-analysis, we applied TEAM to a combination of two functional markers, including (TNF-$\alpha$, IL2), (TNF-$\alpha$, IFN-$\gamma$), (TNF-$\alpha$, CD107), (IL2, IFN-$\gamma$), (IL-2, CD107), and (IFN-$\gamma$, CD107). In each sub-analysis, we used the sequential partitioning algorithm with 148 bins along each dimension (total $148^2 = 21,904$ bins, with $n \approx 210$ cells in each bin), a maximum of three layers, and a nominal FDR of 0.05. As in our simulations, the choice of $n$ was set such that $n \approx \{2(N_1 + N_2)\}^{1/3}$.

Each sub-analysis identified the differential PDF bins, and we can find the cells that are located in those identified bins. If the T cells are activated in one functional marker (monofunctional), we expect them to fall in the differential region in three sub-analyses involving this functional marker. Similarly, if the T cells are activated in two (bifunctional) or three functional markers (polyfunctional), they are expected to fall in the differential region in five or six sub-analyses.

Based on the biological classification, we classify the cohort 2 cells into four groups: nonfunctional (falling into the identified regions in fewer than two sub-analyses: 97.9% cohort 2 cells), monofunctional (falling into the identified region in three or four sub-analyses: 1.8% of cohort 2 cells), bifunctional (falling into the identified region in five sub-analyses: 0.16% of cohort 2 cells), and polyfunctional (falling into the identified region in six sub-analyses: 0.23% of cohort 2 cells).

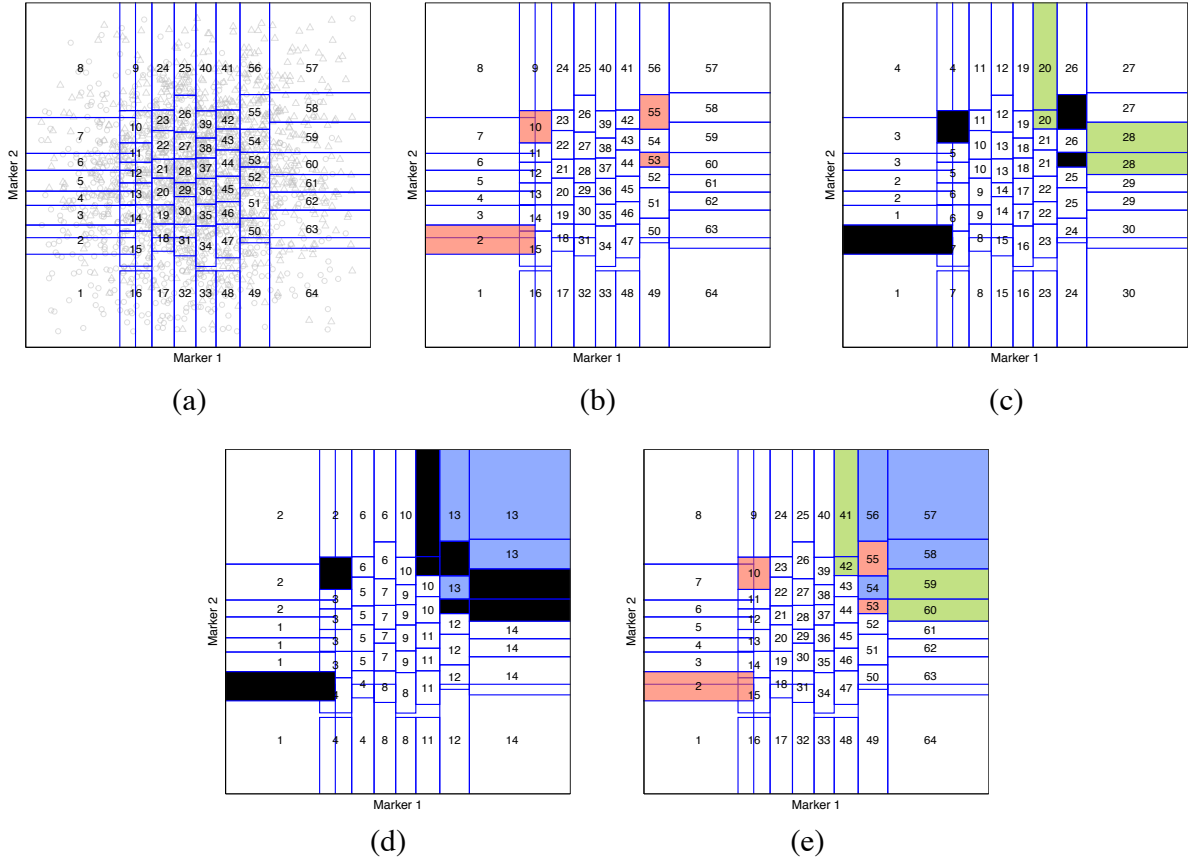

Figure 3: An example of sequential partitioning and aggregation in a two-dimensional sample space. We partition the sample space into 64 bins to facilitate visualization. (a) Partition the two-dimensional sample space using the sequential partition strategy. Then all the bins are sequentially ordered. (b) On layer 1, the rejection set is $\mathcal{R}^{(1)} = \{2, 10, 53, 55\}$. (c) On layer 2, the remaining bins are aggregated with their neighbors along the order defined by sequential partition. The indices of each bin in layer 2 are relabeled and comprise the bin set $S_i^{(2)}$. After the testing procedure, null hypotheses coupled with bin set 20 and 28 are rejected. After mapping the rejections to layer 1, the rejection set is $\mathcal{R}^{(2)} = \{41, 42, 59, 60\}$. (d) On layer 3, we further aggregate the neighboring bin sets. After the testing procedure on layer 3, we reject the null hypotheses coupled with bin set 13. After mapping the rejections to layer 1, the rejection set is $\mathcal{R}^{(3)} = \{54, 56, 57, 58\}$. After three layers, the overall rejection set is $\mathcal{R} = \{2, 10, 53, 54, 55, 56, 57, 58, 59, 60\}$.

The 1D functional protein marker densities are plotted in Supplementary Figure S1. All functional markers are more highly expressed in monofunctional, bifunctional, and polyfunctional T cells, in agreement with our intuitions. The 2D densities of the lineage (Figure 4A) and maturational (Figure 4B) protein marker expressions are plotted to evaluate the distributions of the monofunctional, bifunctional, polyfunctional T cells against the nonfunctional T cells. In Figure 4A, CD3 is a T cell co-receptor, and CD4 and CD8 are cell surface proteins that characterize T cells in helper and cytotoxic subtypes, respectively. The first row of Figure 4A shows that CD3 is increasingly down-regulated in more highly activated cells. This is consistent with the internalization and down-regulation of the TCR upon TCR-ligand activation (Valitutti et al., 1997).

The "dump channel" multiplexed markers CD14/CD19/vAmine that share the same fluorochrome are used to exclude monocytes, macrophages, B cells and dead cells. The second row of Figure 4A shows the enrichment of CD4+CD8+ double positive T cells among the polyfunctional cells compared to the nonfunctional cells. The CD4-CD8+ T cells are enriched among either monofunctional or polyfunctional cells, but not apparent among bifunctional cells. In Figure 4B, CD27, CD57, and CD45RO are cell-surface proteins that are differentially expressed in naive, effector and memory T cells. We excluded CD4-CD8- double negative T cells to highlight only the well-known T cells. The first row of Figure 4B shows that the activated T cells are mostly naïve (N) and central memory (CM) T cells, with some polyfunctional cells having an effector memory (EM) phenotype. The second row of Figure 4 shows that the terminal effect (TE) and effector (E) T cells are more enriched in polyfunctional T cells, compared to the monofunctional and bifunctional T cells.

*4.3. Analysis Using Existing Methods*

As a comparison, we applied the scan-statistic method in Walther et al. (2010) and MRS in Soriano and Ma (2017) to the same data. Unfortunately, the scan-statistic method in Walther et al. (2010) failed to complete the analysis within one week on a 2.10GHz Intel Xeon Gold 6252 CPU laptop, while TEAM only took 15 minutes. Each run of MRS up to six (default) or seven layers takes about 5 hours. We also ran MRS using the maximum-allowable tree depth of 14, in order to match the same level of high-resolution as the first layer of TEAM($\log_2(21,904) \approx 14$). However, this analysis did not complete within one week on the same machine.

We applied MRS to our dataset under two settings: 1) using only information in the four functional markers, TNF-$\alpha$, IL-2, IFN-$\gamma$, and CD107, and 2) using the information on all 11 markers. We run MRS up to layer six (default) and seven and controlled the FDR of the discovered bins (same as "windows" defined in Soriano and Ma (2017)) at the 0.05 level.

By using only the four functional markers, MRS did not identify any bins harboring significantly differential density, regardless of the tree depth. By using all eleven markers, we identified several significant bins. Figure 5 displays the maximum *a posteriori* (MAP) tree with depths of six and seven. Red nodes marked by red squares (henceforth called MRS-significant bins) correspond to the significant bins where cohort 2 cells are more enriched.

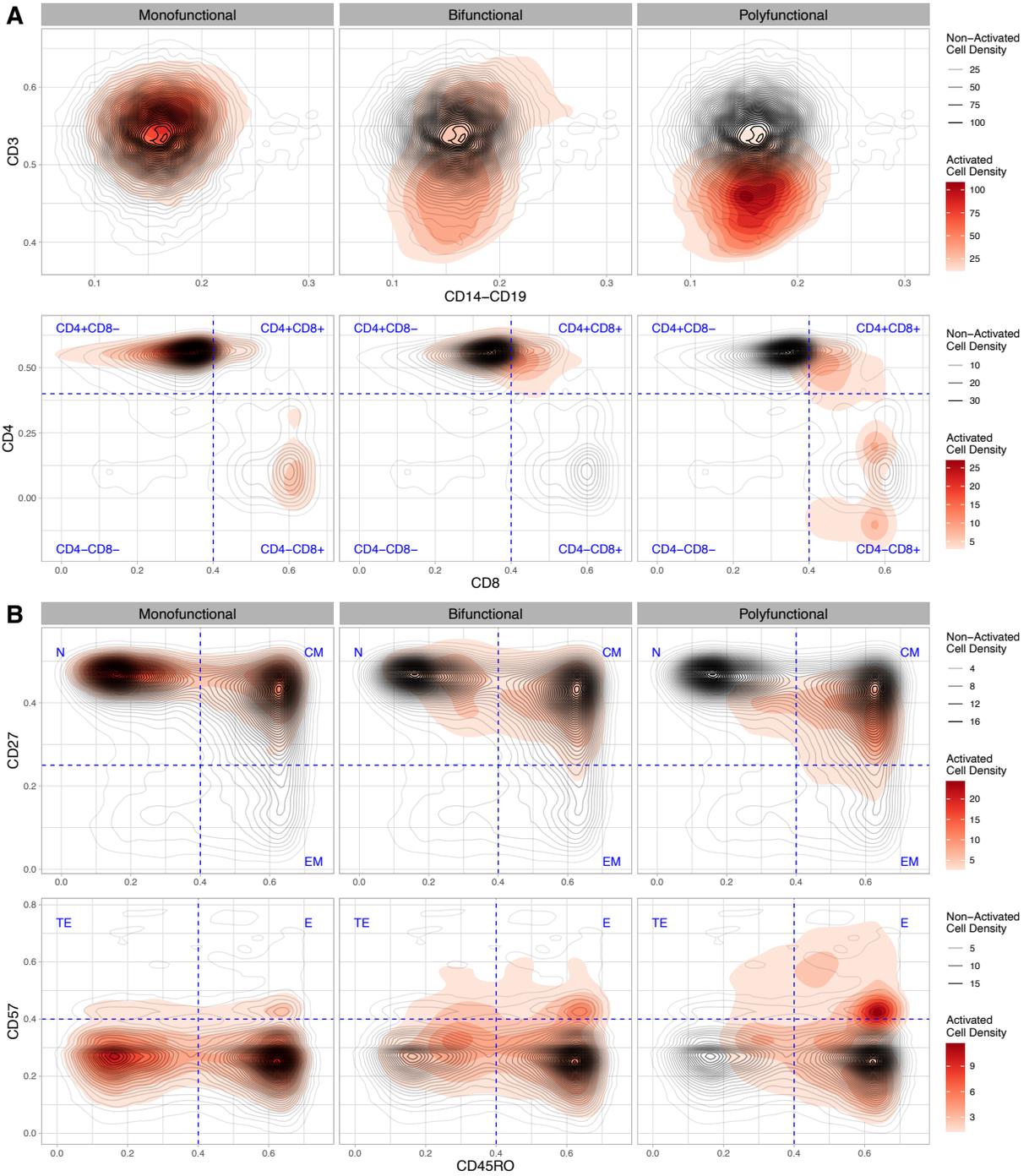

Figure 4: 2D lineage (panel A) and maturational (panel B) marker densities of the monofunctional, bifunctional, and polyfunctional cells against the nonfunctional cells. The cells are classified based on whether and how many times they fall into the TEAM-identified regions. The monofunctional, bifunctional, and polyfunctional cell marker densities are displayed as grayscale contour lines, and the non-functional cell marker densities are displayed as filled, red contours. In panel B, we excluded CD4-CD8- double-negative T cells to focus on the well-known T cell subsets, which include the five respective maturational subsets (N: naïve, CM: central memory, EM: effector memory, TE: terminal effector, E: effector).

We also discovered that most bins identified by a six-layer MRS algorithm and by a seven-layer MRS algorithm are non-overlapping. The Jaccard similarity index (*i.e.* the ratio of intersection to the union of cells in any two bins) for cohort 2 cells is zero across all MRS-significant bins, except for those corresponding to nodes B6 and C7, which have a similarity index of 0.97. The discrepancy in these results is likely due to the different generative priors under each tree depth. As a result, MRS results can be highly variable with different tree depths. This introduces difficulty in interpreting the results. In contrast, each layer of TEAM is fully nested so the region identified by the six-layer TEAM must be the subsets of those identified by the seven-layer TEAM. Computationally, this also saves time by making the analysis additive: to add an extra layer of analysis, start from the existing results and run only one additional layer.

## 5. Numerical Experiments

To demonstrate the performance of TEAM, we designed four simulation settings that mimic the flow cytometry analysis. Settings S1-S3 simulate monofunctional cells that will express one cytokine after being challenged with antigen. As shown in Section 4, the distribution of monofunctional cells is usually the most similar to the distribution of the non-functional cells, and these distributions are the most challenging to be differentiated. In practice TEAM would be applied to compare the multivariate protein marker PDFs in each sub-analysis. To avoid the arbitrariness in simulating another irrelevant protein marker for monofunctional cells, we applied TEAM to compare univariate protein marker distribution differences. To identify monofunctional cells, the only difference between a univariate TEAM analysis and a bivariate TEAM analysis is in how TEAM partitions the sample space. However, if we use the sequential partition and aggregation as illustrated in Figure 3, the results of the univariate TEAM and the bivariate TEAM are similar. In setting S4, we simulate the bifunctional cells that are activated in two protein markers and applied the bivariate TEAM analysis to this setting.

The details of settings S1-S4 are listed below. Here $Y$ represents the cellular protein marker expression. Under all settings, we simulate only a very small proportion of cohort 2 cells to be activated, so that the cohort 2 cell density only deviates from the cohort 1 cell density a little to increase the challenge. This also mimics real flow cytometry studies, because only a small proportion of cells are expected to be activated (as illustrated in Section 4). Figure 6 illustrates the protein marker distributions of two cohorts of cells under settings S1-S4.

S1. **One-dimensional local shift.** We randomly generated $N_1 = N_2 = 1,474,560$ random samples in each cohort. Assume

$$Y \sim \begin{cases} 0.97\mathcal{N}(0.4, 0.04^2) + 0.03\mathcal{N}(0.88, 0.01^2) & \text{in cohort 1} \\ 0.97\mathcal{N}(0.4, 0.04^2) + 0.03\mathcal{N}(0.89, 0.01^2) & \text{in cohort 2}. \end{cases}$$

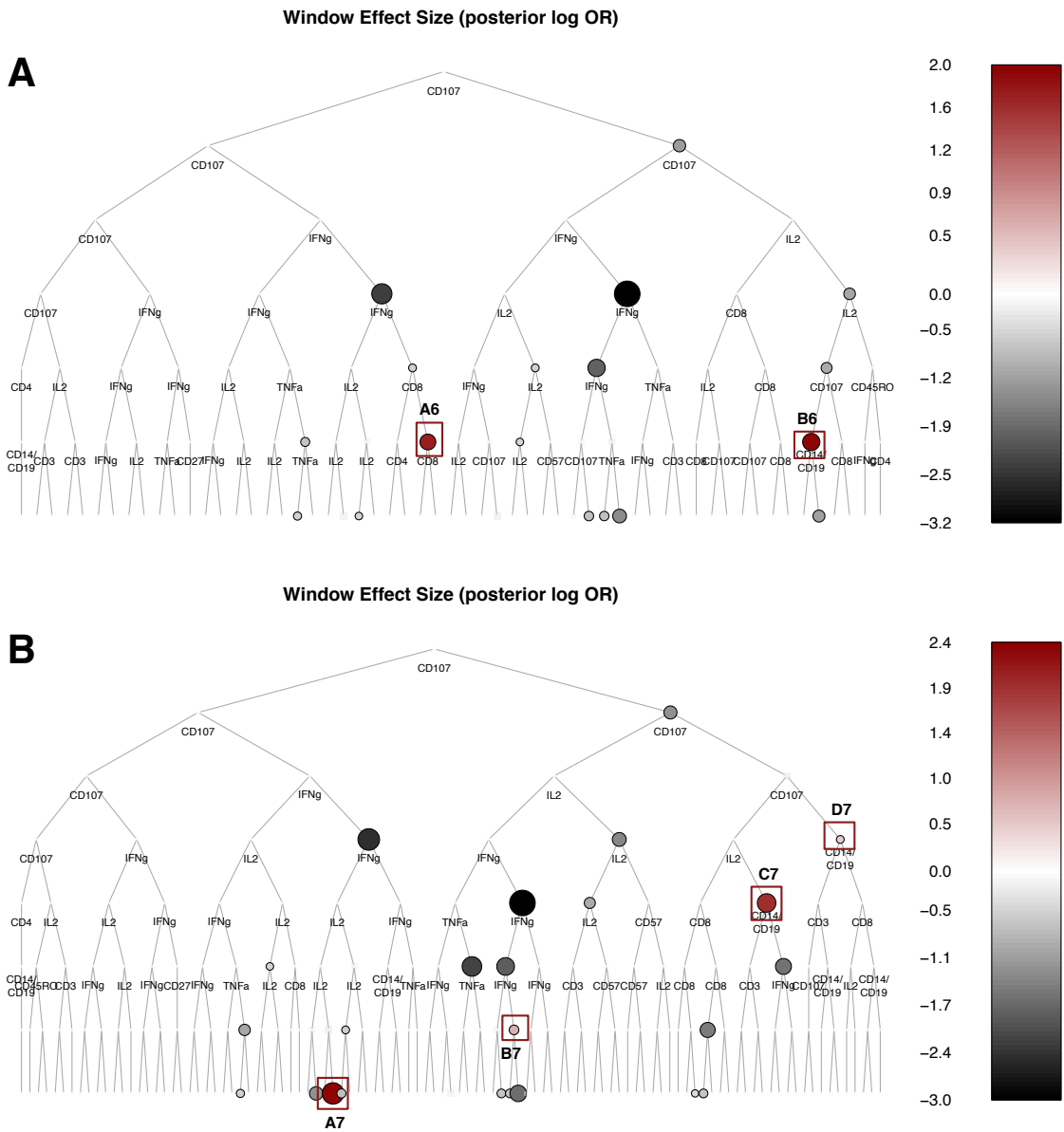

Figure 5: The MAP partition trees from the MRS algorithm applied to the EQAPOL dataset with A) tree depth of 6 and B) tree depth of 7. All 11 protein markers were used in the analyses. The label underneath each node (represented by a circle) indicates the marker/dimension being partitioned to get the two children. Nodes in red and black represent windows with positive and negative log odds ratios (log OR), respectively. The size of the node corresponds to the magnitude of the log OR. The red rectangles (*e.g.* A6 and B6 in A) highlight the regions with positive log OR and claimed to harbor differential density by MRS.

S2. **One-dimensional local dispersion difference.** We randomly generated $N_1 = N_2 = 1,474,560$ random samples in each cohort. Assume

$$Y \sim \begin{cases} 0.97\mathcal{N}(0.4, 0.04^2) + 0.03\mathcal{N}(0.8, 0.02^2) & \text{in cohort 1} \\ 0.97\mathcal{N}(0.4, 0.04^2) + 0.03\mathcal{N}(0.8, 0.03^2) & \text{in cohort 2.} \end{cases}$$

S3. **One-dimensional local shift plus dispersion difference.** We randomly generated $N_1 = N_2 = 1,474,560$ random samples in each cohort. Assume

$$Y \sim \begin{cases} 0.97\mathcal{N}(0.4, 0.04^2) + 0.03\mathcal{N}(0.8, 0.04^2) & \text{in cohort 1} \\ 0.97\mathcal{N}(0.4, 0.04^2) + 0.03\mathcal{N}(0.82, 0.05^2) & \text{in cohort 2.} \end{cases}$$

S4. **Two-dimensional Gaussian mixture with extra cohort 2 samples in some subregions.** We generated $N_1 = N_2 = 500,000$ random samples for each cohort. For all cohort 1 samples and $495,000$ cohort 2 samples, $Y \sim \sum_{j=1}^{5} p_j N(\mu_j, \Sigma_j)$, where

$(p_1, \ldots, p_5) = (0.03, 0.14, 0.17, 0.26, 0.40),$
$\mu_1 = (9,9)', \mu_2 = (3.4, 2.9)', \mu_3 = (8.7, -5.8)', \mu_4 = (-0.4, 3.5)', \mu_5 = (-6, -6.5)',$
$\Sigma_1 = \begin{pmatrix} 2.1 & 0.6 \\ 0.6 & 0.9 \end{pmatrix}, \Sigma_2 = \begin{pmatrix} 0.71 & 0.14 \\ 0.14 & 2.12 \end{pmatrix}, \Sigma_3 = \begin{pmatrix} 2.08 & 0.87 \\ 0.87 & 1.39 \end{pmatrix}, \Sigma_4 = \begin{pmatrix} 2.8 & 0.6 \\ 0.6 & 1.2 \end{pmatrix}, \Sigma_5 = \begin{pmatrix} 1.34 & -0.45 \\ -0.45 & 3.13 \end{pmatrix}.$

For cohort 2, we generated an additional $5,000$ activated samples (1% of cohort 2), $Y$ follows the uniform distribution in the $\mathbb{R}^2$ subregion

$$\{(6, 8.5) \otimes (9, 11)\} \cup \{(8, 10) \otimes (7, 8)\} \cup \{(10, 11) \otimes (9, 12)\}.$$

Under settings S1-S3, we applied the one-dimensional sequential partition algorithm to obtain $m^{(1)} = 2^{14}$ bins. Each bin contains about $n = 180$ pooled samples. The number $n$ is set such that $n \approx \{2(N_1 + N_2)\}^{1/3}$. We tried other choices of $n \in [80, 250]$, and the results of TEAM were similar. The number of layers $L$ was chosen so that $1000 \leq m^{(L)} < 2000$. In settings S1-S3, $m^{(5)} \approx 2^{10} = 1024$. Thus, we ran TEAM up to five layers. We also tried to run TEAM up to layer 6, but many times TEAM did not reject any additional hypotheses on layer 6, suggesting that running five layers might be sufficient. Under setting S4, we applied the two-dimensional sequential partition algorithm to obtain $m^{(1)} = 90^2 = 8100$ bins. Each bin has about 120 samples. Based on the same $L$ selection criterion, we ran TEAM up to layer 4.

Each experiment was repeated 1000 times. TEAM is very computationally efficient. One repetition only takes less than 5 seconds on a 2.6 GHz Intel Core i5 processor with 16Gb memory. The code for running TEAM is available at https://github.com/jbp7/TEAM.

Figure 7 shows how TEAM performs when stopping at layer $L = 1, 2, 3, 4, 5$. Specifically, it shows the average realized false discovery proportion, the average number of false negatives, and the average total discoveries under the four settings. TEAM successfully controls the FDR under the desired level after 5 layers. Meanwhile, the average false negatives substantially goes down as TEAM proceeds to higher layers. For example, under setting S1, at the desired FDR level 0.05, the single layer method misses more than 80 alternative bins (out of about 245 total alternative bins); a five-layer TEAM only misses about 40 alternative bins. The power substantially increases as the layer goes up. Clearly, TEAM will make more true rejections on higher layers.

We compared the performance of TEAM with the MRS method proposed by Soriano and Ma (2017). When we ran MRS, we set the MRS-type FDR at the levels 0.05, 0.10, 0.15, and 0.20. It is important to clarify that this MRS-type FDR is not the same as the FDR defined on those hypotheses on the finest resolution level. See explanations in Section 1. Because MRS is a two-sided testing method embedded in the partition tree, we perform the following procedure to obtain comparable effective discoveries and conduct a fair comparison. First, we ran MRS up to 14 layers under settings S1-S3 and 13 layers under setting S4. Second, we summarized discoveries made in the last 5 layers under settings S1-S3 and the last 4 layers under setting S4. This was to make sure that the layers we compared in TEAM and MRS had comparable resolution levels. Third, among those discoveries, we summarized those with effect size favoring cohort 2. This is because MRS performs a two-sided test while TEAM performs a one-sided test. Fourth, we mapped those discoveries to the finest resolution layer (the last layer). We call those discovered bins the "effective" discoveries. This is to mimic the mapping step in TEAM to calculate the realized FDP on the finest layer.

In Table 1, we report the percentage of repetitions where MRS made at least one rejection, and among those repetitions with one or more rejections, the average number of rejections and the realized FDP when mapping discoveries to the finest resolution layer. This realized FDP has the same definition as the TEAM algorithm. It is easy to see that at fine-resolution levels, the performance of MRS is not satisfying. Under settings S1-S3, MRS fails to identify any bins in the last 5 fine-resolution layers in most repetitions. This is not surprising because we simulated very challenging situations. However, these situations are common in flow cytometry analysis. Among those repetitions with identified bins, the realized FDP is also high. Under setting S4, MRS can identify bins in the last 5 fine resolution layers in most repetitions, but the corresponding realized FDP is extremely high. The fundamental reason for MRS's poor performance under these settings is because MRS is a tool designed for visualizing global multi-dimensional distribution differentiation and automatically searching for splitting direction; it is not designed for pinpointing fine-resolution differential density regions. Thus, while MRS has its own advantages and will be appropriate for certain application scenarios, it is not suitable for pinpointing the small density differential regions that harbor the activated cells.

## 6. Discussion

In this paper, we proposed a novel method to pinpoint the activated T cells in a flow cytometry experiment after stimulation with an antigen. Statistically, we translated this problem to pinpoint the differential density regions using a multiple-testing procedure on an aggregation tree. In practice, TEAM can be applied to two- or three- dimensional protein markers in one sub-analysis and the results of all sub-analyses can be combined to identify the activated monofunctional, bifunctional, and polyfunctional cells. Following this pipeline, we applied TEAM to a flow cytometry study to identify antigen-specific activated T cells. We successfully identified monofunctional, bifunctional, and polyfunctional T cells and characterized their activation in the context of enriched T cell subtypes, while other competing methods fail to provide such comprehensive information.

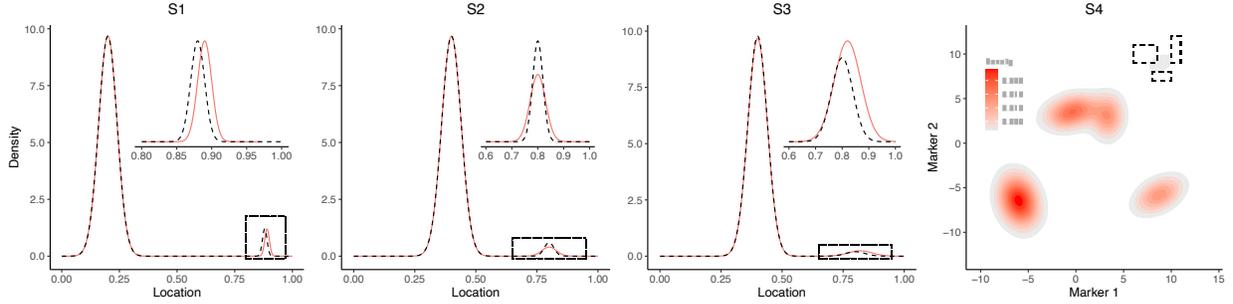

Figure 6: Density functions under S1 - S4. In the first three panels, the red curve represents the PDF of cohort 1 and the dashed curve represents PDF of cohort 2. In the fourth panel, the Gaussian mixture PDF of cohort 1 is illustrated by the heat-map, and the dashed rectangle illustrates the uniform distribution regions of the 1% activated samples in cohort 2.

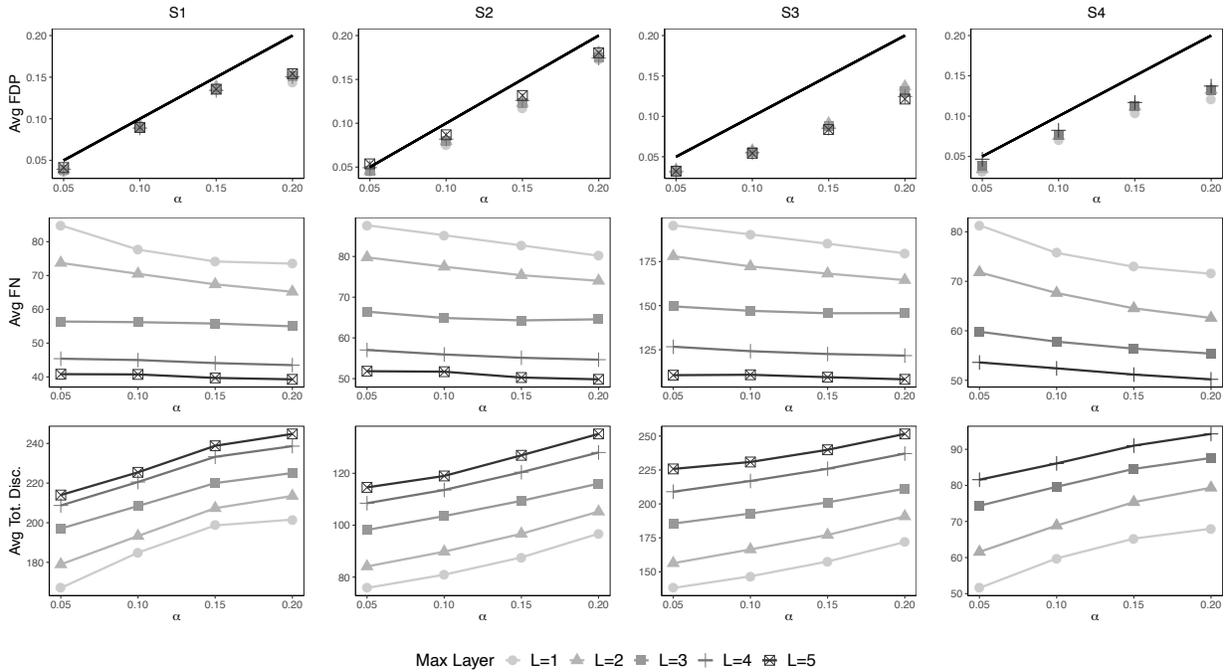

Figure 7: Performance of TEAM under S1 - S4. The first row shows the average realized false discovery proportion with TEAM stopping at different layers ($L = 1, 2, 3, 4, 5$). The second row shows the average number of false negatives when TEAM stops at different layers. As a reference, the average numbers of alternative leaves under S1 - S4 are 245, 160, 329, 130. The third row shows the average number of total discoveries of TEAM when stopping at different layers.

|  | MRS-type FDR level | | | | | | | | | | | |
|---|---|---|---|---|---|---|---|---|---|---|---|---|
|  | 0.05 | | | 0.10 | | | 0.15 | | | 0.20 | | |
|  | % Rej | Avg Rej | Avg FDP | % Rej | Avg Rej | Avg FDP | % Rej | Avg Rej | Avg FDP | % Rej | Avg Rej | Avg FDP |
| S1 | 0.8 | 10 | 0.52 | 3.0 | 8.5 | 0.44 | 9.8 | 980.5 | 0.54 | 25.6 | 3999.9 | 0.63 |
| S2 | 0.9 | 1.1 | 0.11 | 1.4 | 1.3 | 0.14 | 2.7 | 1.5 | 0.16 | 5.4 | 1.4 | 0.22 |
| S3 | 1.9 | 1.3 | 0.16 | 3.3 | 1.4 | 0.17 | 7.6 | 629 | 0.23 | 16.4 | 2147.2 | 0.31 |
| S4 | 95.8 | 297.7 | 0.69 | 96.9 | 285.8 | 0.68 | 97.3 | 252.7 | 0.68 | 98.2 | 238.9 | 0.67 |

Table 1: Performance of MRS in S1-S4. Each setting has 1000 repetitions. "% Rej" is percentage of repetitions where MRS made at least one effective discoveries; "Avg. Rej" is, among those repetitions with one or more effective discoveries, the average number of discovered bins when MRS's effective discoveries are mapped to the finest resolution layer; "Avg. FDP" is, among those repetitions with one or more effective discoveries, the average realized false discovery proportion after MRS's effective rejection are mapped to the finest resolution layer.

Although we focused on the application on flow cytometry analysis in this paper, TEAM can be used to draw insights from data in other contexts. For example, TEAM can also be applied to single-cell RNA sequencing (scRNA-seq) data to identify the activated cells under stimulation. The current limitation of TEAM is that it can only work on two or three markers at a time due to the exponential growth in number of bins, as dimensionality increases. Thus, dimensional-reduction methods need to be applied to high-dimensional data (such as scRNA-seq data) first, or signature markers need to be extracted. In the future, we will further study new methods that can be directly applied to identify the activated cells with high-dimensional markers.

## Acknowledgements

Flow cytometric data generation was supported in whole or part through an EQAPOL collaboration with federal funds from the National Institute of Allergy and Infectious Diseases, National Institutes of Health, Contract Number HHSN272201700061C.

John Pura's research was supported by NSF DGE 1545220 and the NIH training grant T32HL079896. Cliburn Chan's research was supported by EQAPOL Contract Number HHSN272201700061C, and the Duke University Center for AIDS Research (CFAR), and NIH funded program (5P30 AI064518). Jichun Xie's research was supported by the Duke Cancer Institute and the Duke Center of Human System Immunology.

# Supplemental Materials of "TEAM: A Multiple Testing Algorithm on the Aggregation Tree for Flow Cytometry Analysis"

In the supplementary materials, we provide the theoretical justification for TEAM and additional plots referenced in the main text.

## S1. Justification of the aggressive rejection rule

On layer $\ell$, after we aggregated the leaf bins into larger bins, for their index set $S_i^{(\ell)}$ ($i \in \{1, \ldots, m^{(\ell)}\}$), we define

$$\mathrm{H}_{\mathrm{nul},i}^{(\ell)} : \ \forall \ j \in S_i^{(\ell)}, \theta_j \leq \theta_0 \quad \text{versus} \quad \mathrm{H}_{\mathrm{alt},i}^{(\ell)} : \ \exists \ j \in S_i^{(\ell)}, \theta_j > \theta_0. \tag{S1}$$

When $\mathrm{H}_i^{(\ell)}$ is rejected, all $\mathrm{H}_{\mathrm{nul},j}$ with $j \in S_i^{(\ell)}$ are rejected. This rejection rule is very aggressive. Yet, it makes sense under the assumption that null (alternative) hypotheses tend to cluster together along the order. In other words, if $\mathrm{H}_{\mathrm{alt},i}$ is true, then its neighboring hypotheses are more likely to be alternative. This underlying assumption is reasonable in the setting of identifying differential density regions especially when both $f_1$ and $f_2$ satisfy the following regularity conditions.

**Lemma 2.** *Suppose that $f_1(y)$ and $f_2(y)$ are two probability density functions on a closed sample space $\Omega \subset \mathbb{R}^p$ satisfying*

$$0 < M_1 \leq \inf_{s \in \{1,2\}, \ y \in \Omega} f_s(y) \leq \sup_{s \in \{1,2\}, \ y \in \Omega} f_s(y) \leq M_2,$$

*and*

$$\sup_{s \in \{1,2\}, \ y \in \Omega} \|\nabla f_s\|_2 \leq M_3, \quad \text{where } \nabla f_s = \left(\frac{\partial f}{\partial y_1}, \ldots, \frac{\partial f}{\partial y_p}\right)'.$$

*Assume the partition $\{\Omega_1, \ldots, \Omega_m\}$ on $\Omega$ satisfies*

$$\sup_{i=1}^{m} \mu(\Omega_i) \leq M_4/m, \quad \text{and} \quad \sup_{i=1}^{m} \mathsf{Span}(\Omega_i) \leq M_5/m^{1/p}.$$

*For some constant $M_4, M_5 > 0$. Here $\mu$ is the Lebesgue measure and $\mathsf{Span}(\Omega_i) = \sup_{y_1, y_2 \in \Omega_i} \|y_1 - y_2\|_2$. Then*

$$\sup_{i \in \{2, \ldots, m\}} |\theta_i - \theta_{i-1}| \leq O(1/m). \tag{S2}$$

Lemma 2 requires several mild conditions. The first condition requires that $f_1$ and $f_2$ are bounded functions with bounded first derivatives. This condition holds for a wide range of functions. The second condition requires the partition to be roughly "even" on the sample space $\Omega$ and across all dimensions. The second condition is satisfied with many partition methods, for example, the sequential partition strategy.

**Proposition 1.** *With the probability converging to 1, the partition $\{\Omega_1, \ldots, \Omega_m\}$ generated based on sequential partition strategy satisfies*

$$\sup_{i=1}^{m} \mu(\Omega_i) \leq M_4/m, \quad \text{and} \quad \sup_{i=1}^{m} \mathsf{Span}(\Omega_i) \leq M_5/m^{1/p}.$$

*For some constant $M_4, M_5 > 0$.*

## S2. Asymptotic Validity

To simplify, we only consider TEAM to stop at layer $L$, where $L$ is a constant. First, we introduce the conditions needed to justify the theoretical properties of TEAM. To unify the expression of the conditions on all layers $l \in \{1, \ldots, L\}$, let $\mathcal{B}^{(\ell)} = \{1, \ldots, m^{(\ell)}\}$, the null set $\mathcal{B}_{\text{nul}}^{(\ell)} = \{i : \forall\, j \in S_i^{(\ell)},\, \theta_j \leq \theta_0\}$, and the alternative set $\mathcal{B}_{\text{alt}}^{(\ell)} = \mathcal{B}^{(\ell)} \setminus \mathcal{B}_{\text{nul}}^{(\ell)}$. Also let $m_0^{(\ell)} = \mathsf{Card}(\mathcal{B}_{\text{nul}}^{(\ell)})$, $m_1^{(\ell)} = \mathsf{Card}(\mathcal{B}_{\text{alt}}^{(\ell)})$. Then $m^{(\ell)} = m_0^{(\ell)} + m_1^{(\ell)}$.

C1. Assume $m_1^{(1)} \leq r_2 \{m^{(1)}\}^{r_1}$ for some $r_1 < 1/2$, and $r_2 > 0$. Let $n^{(1)} = \sup_{i \in \{1,\ldots,m\}} n_i$. Assume $n_i \geq n^{(1)}[1 - o(\sqrt{\frac{1}{n \log m}})]$ for all $i \in \{1,\ldots,m\}$ and $N^{r_3} \leq n^{(1)} \leq N^{r_4}$ for some constants $\frac{r_1}{p+r_1} < r_3 \leq r_4 < \frac{1-r_1}{1+p-r_1}$, where $N = N_2 + N_1$ is the total number of cells.

C2. For all $i \in \{1,\ldots,m\}$, assume $r_5 \leq \theta_i \leq 1 - r_5$ for some constants $r_5$ satisfying $0 < r_5 < 0.5$.

C3. Let $\alpha^{(0)} = +\infty$. For any $1 \leq \ell \leq L$, let

$$\beta = \left\{2(1-r_1)\log m^{(1)} - \log\log m^{(1)}\right\}^{1/2},\ \gamma = (2\log m^{(1)})^{1/2},\ \lambda = (2\log\log m^{(1)})^{1/2} \tag{S3}$$

Define

$$\mathcal{G}^{(1)} = \left\{i : (n^{(1)})^{1/2}(\theta_i - \theta_0) > 2^{-1}\lambda + \{\theta_0(1-\theta_0)\}^{1/2}\gamma\right\}.$$

For $\ell \geq 2$, define

$$\mathcal{E}_i^{(\ell)} = \{i, i+1, \ldots, i+2^\ell - 3\} \subseteq \{1, \ldots, m\},$$

$$\mathcal{G}^{(\ell)} = \Big\{i \in \{1,\ldots,m^{(1)}\} : \forall\, j \in \mathcal{E}_i^{(\ell)},$$

$$2^{-1}\lambda + \{\theta_0(1-\theta_0)\}^{1/2}\gamma < (n^{(\ell)})^{1/2}(\theta_j - \theta_0) \leq 2^{-1/2}\beta\Big\}. \tag{S4}$$

Assume for some constant $r_6 > 0$, $s_G^{(\ell)} = \mathsf{Card}\{\mathcal{G}^{(\ell)}\}$ satisfies

$$s_G^{(\ell)} \geq r_6 \log m^{(1)}.$$

C4. The neighboring $\theta_i$s are similar such that

$$\sup_{j=2,\ldots,m} |\theta_j - \theta_{j-1}| = o\left(\frac{1}{n^{(1)}\log m^{(1)}}\right).$$

Condition C1 assumes that the true alternative bins are sparse. The condition $r_1 < 1/2$ is sufficient to guarantee that in (S4)

$$2^{-1}\lambda + \{\theta_0(1-\theta_0)\}^{1/2}\gamma < \sqrt{2\theta_0(1-\theta_0)}\beta.$$

Condition C1 also specifies a lower bound for the number of the pooled cells in each bin. This number cannot be too small to affect the asymptotic convergence in the individual bin test. Also, because $m^{(1)} = N/n^{(1)}$, it also imposes a upper bound for the number of hypotheses on the bottom layer. The condition $N^{r_3} \leq n^{(1)} \leq N^{r_4}$ is equivalent to the condition $(n^{(1)})^{\frac{1-r_4}{r_4}} \leq m^{(1)} \leq (n^{(1)})^{\frac{1-r_3}{r_3}}$, where $m^{(1)}$ is the number of hypotheses on the bottom layer. Besides, $n_i \geq n^{(1)}[1 - o(\sqrt{\frac{1}{n \log m}})]$ requires the samples in each bin are almost the same. Both the *adaptive partition* and the *sequential partition* described in Section 3 satisfy this condition.

Condition C2 assumes all $\theta_i$ is bounded away from 0 and 1.

Condition C3 assumed the existence of clustered signals with certain strength levels. The corresponding signal sets are labelled by $\mathcal{G}^{(1)}, \ldots, \mathcal{G}^{(L)}$. Those $\mathcal{G}^{(\ell)}$ with smaller $\ell$ contain signal segments where the individual signal is strong ($\theta_i$ is large) but the segment is short; Those $\mathcal{G}^{(\ell)}$ with larger $\ell$ contain signal segments where the individual signal is weak ($\theta_i$ is close to $\theta_0$) but the segment is long. The signal level and segment length in $\mathcal{G}^{(\ell)}$ is designed in a way such that at least a sub-segment with length $2^{\ell-1}$ will stay after $\ell - 1$ layers with a non-negligible probability; and on layer $\ell$, this sub-segment will be identified with a high probability converging to 1. The cardinality condition $s_G^{(\ell)} \geq r_6 \log m^{(1)}$ is for the simplicity of presentation in proof. In fact, we only require $\lim_{N \to +\infty} s_G^{(\ell)} = +\infty$. Compared with the upper bound on the total alternatives $r_2(m^{(1)})^{r_1}$, $s_G^{(\ell)}$ is much smaller. This means that we only require some small numbers of signals to be large.

Conditions C1–C3 are required to prove that the layer-specific FDP of TEAM is consistent. Condition C4 is required to prove the overall FDP is consistent. Lemma 2 outlines the condition needed to guarantee

$$\sup_{i \in \{2,\ldots,m\}} |\theta_i - \theta_{i-1}| \leq O(1/m^{(1)}).$$

By Condition C1, $m^{(1)} \geq \{n^{(1)}\}^{\frac{1-r_4}{r_4}} > \{n^{(1)}\}^{2/3}$. Therefore, $O(1/m^{(1)}) = o\left(\frac{1}{n^{(1)} \log m^{(1)}}\right)$ satisfying Condition C4.

Recall our definitions on the parent null and alternative hypothesis sets

$$\mathcal{B}_{\text{nul}}^{(\ell)} = \{i : S_i^{(\ell)} \subseteq \mathcal{H}^{(\ell)}, \forall j \in S_i^{(\ell)}, \theta_j \leq \theta_0\}, \quad \mathcal{B}_{\text{alt}}^{(\ell)} = \{i : S_i^{(\ell)} \subseteq \mathcal{H}^{(\ell)}, \exists j \in S_i^{(\ell)}, \theta_j > \theta_0\}.$$

Based on them, we define the false and true rejection sets

$$\mathcal{V}^{(\ell)} = \{i \in \mathcal{B}_{\text{nul}}^{(\ell)} : S_i^{(\ell)} \subseteq \mathcal{R}^{(\ell)}\}, \quad \mathcal{W}^{(\ell)} = \{i \in \mathcal{B}_{\text{alt}}^{(\ell)} : S_i^{(\ell)} \subseteq \mathcal{R}^{(\ell)}\}.$$

Also define $\mathcal{U}^{(\ell)} = \mathcal{V}^{(\ell)} \cup \mathcal{W}^{(\ell)}$. Let

$$U^{(\ell)} = \mathsf{Card}(\mathcal{U}^{(\ell)}), \quad V^{(\ell)} = \mathsf{Card}(\mathcal{V}^{(\ell)}), \quad W^{(\ell)} = \mathsf{Card}(\mathcal{W}^{(\ell)}).$$

On layer $\ell$, the layer-specific FDP is defined as $FDP^{(\ell)} = V^{(\ell)}/U^{(\ell)}$.

We will first prove that $FDP^{(\ell)}$ converges to the desired level $\alpha$ in probability. There are two major technical difficulties. First, for any $i \in \mathcal{B}_{\text{nul}}^{(\ell)}$, we know that $\theta_j \leq \theta_0$. However, we calculate the P-value $P_{0,i}^{(\ell)}$ of $X_i^{(\ell)}$ by assuming $\theta_j = \theta_0$. If lots of $\theta_j$s are much smaller than $\theta_0$, many $P_{0,i}^{(\ell)}$ will be too conservative. Fortunately, the following lemma guarantees that this will not happen.

**Lemma 3.** *Define the set*

$$\mathcal{D}_{\text{nul}}^{(1)} = \{i \in \mathcal{H}_{\text{nul}}^{(1)} : \theta_0 - (n^{(1)} \log m^{(1)})^{-1} < \theta_i \leq \theta_0\} \tag{S5}$$

*Then under Condition C1,*

$$\lim_{N \to \infty} \frac{\text{Card}(\mathcal{D}_{\text{nul}}^{(1)})}{m_0^{(1)}} = 1.$$

Now we proceed to prove the validity of TEAM. For $\ell = 1$, the proof is similar to those in the existing references on single-layer multiple testing, *e.g.*, Liu et al. (2014), Xie and Li (2018), and Xia et al. (2019). For higher layer $\ell \geq 2$, because the aggregation and testing on layer $\ell$ depends on the testing results on layer $\ell - 1$, we need to prove the validity by induction.

**Theorem 1.** *Under Conditions C1-C3, on layer $\ell$, TEAM satisfies*

$$\lim_{N \to \infty} \mathsf{P}(|FDP^{(\ell)} - \alpha| \leq \epsilon) = 1 \text{ for any } \epsilon > 0, \text{ and } \lim_{N \to \infty} FDR^{(\ell)} = \alpha.$$

Theorem 1 shows the layer-specific consistency of $FDP^{(\ell)}$. However, this does not necessarily lead to the overall consistency of $FDP$. This is because for any $i \in \mathcal{B}_{\text{alt}}^{(\ell)}$ it is possible that only one or a few $\theta_j > \theta_0$. On the other hand, TEAM has an aggressive rejection rule – once $\text{H}_{\text{nul}}^{(\ell)}$ is rejected, we will reject all the child null $\text{H}_{\text{nul},j}^{(1)}$ for all $j \in S_i^{(\ell)}$. Without additional condition, this aggressive rejection rule will introduce many false positives. However, with Condition C4 (the similar neighboring $\theta_i$ condition), we show that TEAM will have overall FDP converging to $\alpha$ in probability. Specifically, after mapping the rejections on the layer $\ell$ to the bottom layer, denote by $\mathcal{R}_{\text{nul}}^{(\ell)}$, $\mathcal{R}_{\text{alt}}^{(\ell)}$, and $\mathcal{R}^{(\ell)}$ as follows.

$$\mathcal{R}^{(\ell)} = \{j : i \in \mathcal{U}^{(\ell)}, j \in S_i^{(\ell)}\}, \quad \mathcal{R}_{\text{nul}}^{(\ell)} = \{j : i \in \mathcal{U}^{(\ell)}, j \in S_i^{(\ell)} \cap \mathcal{H}_{\text{nul}}\}, \quad \mathcal{R}_{\text{alt}}^{(\ell)} = \mathcal{R}^{(\ell)} \setminus \mathcal{R}_{\text{nul}}^{(\ell)}.$$

It is easy to see that $\text{Card}(\mathcal{R}^{(\ell)}) = 2^{\ell-1} U^{(\ell)}$.

**Theorem 2.** *Under Conditions C1–C4, the overall false discover rate*

$$FDP^{(1:L)} = \frac{\sum_{\ell=1}^{L} \text{Card}(\mathcal{R}_{\text{nul}}^{(\ell)})}{\sum_{\ell=1}^{L} \text{Card}(\mathcal{R}^{(\ell)})} \text{ converges to } \alpha \text{ in probability.}$$

## S3. Proofs of the Main Results

For simplicity of the proof of the asymptotic properties, let $n_i = n$ for $i = 1, \ldots, m$. Then $n^{(\ell)} = 2^{\ell-1}n$. Recall that $S_i^{(\ell)}$ is the index set coupled with node $i$ on layer $\ell$. Further, denote by $\widetilde{G}_i^{(\ell)}(c^{(\ell)}; c^{(\ell-1)})$ the conditional CCDF of $X_i^{(\ell)}$ conditioning on $X_{i_1}^{(\ell-1)} \leq c^{(\ell-1)}, X_{i_2}^{(\ell-1)} \leq c^{(\ell-1)}$. When $\ell = 1$, $c^{(0)} =: +\infty$. Then $\widetilde{G}_i^{(1)}(c^{(1)}; c^{(0)})$ decays to the marginal complementary CDF of $\text{Binom}(n_i, \theta_i)$.

For two sequences of real numbers $\{a_n\}$ and $\{b_n\}$, write $a_n = O(b_n)$ if there exists a constant $C$ such that $a_n \leq Cb_n$ holds for all sufficiently large $n$, write $a_n = o(b_n)$ if $\lim_{n\to\infty} a_n/b_n = 0$. If $a_n = O(b_n)$ and $b_n = O(a_n)$, then $a_n \asymp b_n$. If $\lim_{n\to\infty} a_n/b_n = 1$, write $a_n \sim b_n$.

To prove the asymptotic properties of TEAM, we need the following lemmas. Lemma provide a binomial local limit theorem and moderate deviation result for the binomial tail probability. The proofs of Lemmas 4 (a) and 5 can be found in Chapter 8 of Lesigne (2005). The proofs of Lemmas 3 – 11 can be found in Section S4.

**Lemma 4.** *For a $\text{Binom}(n, \theta)$ random variable $\breve{X}$, $0 \leq \breve{x} \leq n$, for any $k$ satisfies $|k - n\theta| < c_n n^{2/3}$ with $\lim_{n\to} c_n = 0$, we have*

a)
$$\mathsf{P}(\breve{X} = k) = \frac{1}{\sqrt{2\pi n\theta(1-\theta)}} \exp\left(-\frac{(\breve{x} - n\theta)^2}{2n\theta(1-\theta)}\right) \cdot (1 + \epsilon_n(k)).$$

*with*
$$\lim_{n\to\infty} \max_{k:|k-n\theta|<c_n n^{2/3}} |\epsilon_n(k)| = 0$$

b) *Let random variable $\breve{X}' \sim \text{Binom}(n', \theta)$, for any $n'$ satisfies $|n' - n| \leq n\delta_0(m, n)$ with $\delta_0(m, n) = o\left(\sqrt{\frac{1}{n \log m}}\right)$. Then, if $|k - n\theta| \leq C\sqrt{n \log m}$ for some constant $C$, we have*

$$\mathsf{P}(\breve{X}' = k) = \frac{1}{\sqrt{2\pi n\theta(1-\theta)}} \exp\left(-\frac{(\breve{x}' - n\theta)^2}{2n\theta(1-\theta)}\right) \cdot (1 + \epsilon'_n(k)).$$

*with*
$$\lim_{n\to\infty} \max_{k:|k-n\theta|<C\sqrt{n \log m}} |\epsilon'_n(k)| = 0$$

**Lemma 5.** *Let $\breve{X}$ be a $\text{Binom}(n, \theta)$ random variable. Suppose that $\{\tau_n\}$ is a sequence of real numbers such that $\lim_{n\to\infty} \tau_n = +\infty$ and $\lim_{n\to\infty} \tau_n n^{-1/6} = 0$. Then*

$$\mathsf{P}(\breve{X} \geq n\theta + \tau_n \sqrt{n\theta(1-\theta)}) \sim \frac{\varphi(\tau_n)}{\tau_n}.$$

*Here, $\varphi(\cdot)$ is the standard normal density function.*

**Lemma 6.** *Consider the $\mathcal{D}_{\text{nul}}^{(1)}$ defined in (S5). For $c^{(\ell-1)} < 1/2$ and $b_i^{(\ell)} \geq n_i^{(\ell)}\theta_0$,*

$$\lim_{n,m\to\infty} \sup_{S_i^{(\ell)} \subseteq \mathcal{D}_{\text{nul}}^{(1)}} \frac{\left|\widetilde{G}_i^{(\ell)}(b_i^{(\ell)}; c^{(\ell-1)}) - G_{0,i}^{(\ell)}(b_i^{(\ell)}; c^{(\ell-1)})\right|}{G_{0,i}^{(\ell)}(b_i^{(\ell)}; c^{(\ell-1)})} = 0.$$

**Lemma 7.**

$$\sup_{b:b-n_i\theta_0<\sqrt{2n\log m\theta_0(1-\theta_0)}} \left|\frac{G_{0,i}^{(1)}(b + o(\sqrt{\frac{n}{\log m}}); \hat{c}^{(0)})}{G_{0,i}^{(1)}(b; \hat{c}^{(0)})} - 1\right| \to 0$$

**Lemma 8.** *For $c^{(\ell-1)} \leq 1/2$ and $b^{(\ell)} \geq n^{(\ell)}\theta_0$,*

$$\lim_{n,m\to\infty} \sup_{\ell\in\{1,...,L\}} \sup_{i\in\mathcal{B}_{\text{nul}}^{(\ell)}} \left\{\widetilde{G}_i^{(\ell)}(b^{(\ell)}; c^{(\ell-1)}) - G_{0,i}^{(\ell)}(b^{(\ell)}; c^{(\ell-1)})\right\} \leq 0. \tag{S6}$$

$$\lim_{n,m\to\infty} \inf_{\ell\in\{1,...,L\}} \inf_{i\in\mathcal{B}_{\text{alt}}^{(\ell)}} \left\{\widetilde{G}_i^{(\ell)}(b^{(\ell)}; c^{(\ell-1)}) - G_{0,i}^{(\ell)}(b^{(\ell)}; c^{(\ell-1)})\right\} \geq 0. \tag{S7}$$

*Also, define*

$$\widetilde{g}_i^{(\ell)}(b^{(\ell)}; c^{(\ell-1)}) = \mathsf{P}(\check{X}_i^{(\ell)} = b^{(\ell)} \mid G_{0,i_1}^{(\ell-1)}(\check{X}_{i_1}^{(\ell-1)}; \hat{c}^{(\ell-2)}) \geq \hat{c}^{(\ell-1)}, G_{0,i_2}^{(\ell-1)}(\check{X}_{i_2}^{(\ell-1)}; \hat{c}^{(\ell-2)}) \geq \hat{c}^{(\ell-1)})$$

$$g_{0,i}^{(\ell)}(b^{(\ell)}; c^{(\ell-1)}) = \mathsf{P}(\check{X}_i^{(\ell)} = b^{(\ell)} \mid G_{0,i_1}^{(\ell-1)}(\check{X}_{i_1}^{(\ell-1)}; \hat{c}^{(\ell-2)}) \geq \hat{c}^{(\ell-1)}, G_{0,i_2}^{(\ell-1)}(\check{X}_{i_2}^{(\ell-1)}; \hat{c}^{(\ell-2)}) \geq \hat{c}^{(\ell-1)},$$

$$\theta_j = \theta_0, \forall\, j \in S_i^{(\ell)}).$$

*Then*

$$\lim_{n,m\to\infty} \sup_{\ell\in\{1,...,L\}} \sup_{i\in\mathcal{B}_{\text{nul}}^{(\ell)}} \left\{\widetilde{g}_i^{(\ell)}(b^{(\ell)}; c^{(\ell-1)}) - g_{0,i}^{(\ell)}(b^{(\ell)}; c^{(\ell-1)})\right\} \leq 0. \tag{S8}$$

$$\lim_{n,m\to\infty} \inf_{\ell\in\{1,...,L\}} \inf_{i\in\mathcal{B}_{\text{alt}}^{(\ell)}} \left\{\widetilde{g}_i^{(\ell)}(b^{(\ell)}; c^{(\ell-1)}) - g_{0,i}^{(\ell)}(b^{(\ell)}; c^{(\ell-1)})\right\} \geq 0. \tag{S9}$$

**Lemma 9.** *Let $\check{X}_i \sim \text{Binom}(n_i^{(1)}, \theta_0)$, with $i = 1, ..., 2^{\ell-1}$, and $\beta_0 = b_0\sqrt{2(1-r_1)\log m}$, with $b_0 = \sqrt{\frac{\frac{3}{4} - \frac{r_1}{2}}{1-r_1}} \in (\frac{1}{\sqrt{2(1-r_1)}}, 1)$. Then,*

a) *Let $b_k = n^{(k)}\theta_0 + \sqrt{n^{(k)}\theta_0(1-\theta_0)}\beta_0$, we have,*

$$\max_{k=1,...,\ell-1} \sup_{b^{(\ell)}\in[n^{(\ell)}\theta_0, n^{(\ell)}\theta_0+\sqrt{n^{(\ell)}\theta_0(1-\theta_0)}\gamma]} \frac{\mathsf{P}(\sum_{i=1}^{2^{\ell-1}}\check{X}_i > b^{(\ell)}, \sum_{i=1}^{2^{k-1}}\check{X}_i > b_k)}{\mathsf{P}(\sum_{i=1}^{2^{\ell-1}}\check{X}_i > b^{(\ell)})} \to 0 \tag{S10}$$

b) *Let $\hat{b}_i^{(\ell)}$ be the value s.t. $G_{0,i}(\hat{b}_i^{(\ell)}; \hat{c}^{(\ell-1)}) = \hat{c}^{(\ell)}$, and*

$$\hat{\tau}_i^{(\ell)} = \frac{\hat{b}_i^{(\ell)} - n_i^{(\ell)}\theta_0}{\{n_i^{(\ell)}\theta_0(1-\theta_0)\}^{1/2}}. \tag{S11}$$

For $k = 1, ..., \ell - 1$, when $P(\hat{\tau}_i^{(k)} \geq \beta_0) \to 1$, we have

$$|P_{0,i}^{(\ell)} - \breve{P}_{0,i}|/\breve{P}_{0,i} \to 0, \text{ and} \tag{S12}$$

$$|G_{0,i}^{(\ell)}(b_i^{(\ell)}; \hat{c}^{(\ell-1)}) - G_{0,i}^{(\ell)}(b_i^{(\ell)}; \hat{c}^{(0)})|/G_{0,i}^{(\ell)}(b_i^{(\ell)}; \hat{c}^{(0)}) \to 0 \tag{S13}$$

**Lemma 10.** *For every layer $\ell = 1, ..., L$, define the set $\mathcal{X}^{(\ell)}$ based on a sequence $\delta_2(m) = o(1)$ as*

$$\mathcal{X}^{(\ell)} = \left\{ x : \left| \frac{\sum_{i \in \mathcal{B}_{\text{nul}}^{(\ell)}} I(P_{0,i}^{(\ell)} < \hat{c}^{(\ell)}) - \sum_{i \in \mathcal{B}_{\text{nul}}^{(\ell)}} P(P_{0,i}^{(\ell)} < \hat{c}^{(\ell)})}{\left\{ \sum_{i \in \mathcal{B}_{\text{nul}}^{(\ell)}} P(P_{0,i}^{(\ell)} < \hat{c}^{(\ell)}) \right\}} \right| > \delta_2(m) \right\} \tag{S14}$$

*If the FDR control holds on layer $1, ..., \ell - 1$, and for $h = 1, ..., \ell$, there exists constant $C^{(h)}$, s.t.*

$$\mathsf{P}(m_0 \hat{c}^{(h)} \geq C^{(h)} \log m) \to 1, \tag{S15}$$

*then under Conditions C1 and C2, the following three statements hold.*

a) *Exists constant $\delta_2(m) = o(1)$, s.t. $\mathsf{P}(\cap_{h=1}^{\ell} \mathcal{X}^{(h)}) \geq 1 - o(1)$.*

b) *On $\cap_{h=1}^{\ell} \mathcal{X}^{(h)}$, there exists some $C$ such that*

$$\hat{c}^{(\ell)} \leq C(m^{(\ell)})^{r_1 - 1}. \tag{S16}$$

c) *Recall $\beta$ defined in (S3) and $\hat{\tau}_i^{(\ell)}$ defined in (S11). On $\cap_{h=1}^{\ell-1} \mathcal{X}^{(h)}$,*

$$\hat{\tau}_i^{(\ell-1)} \geq \beta(1 + o(1)). \tag{S17}$$

**Lemma 11.** *Under Conditions C1 and C2, on $\mathcal{X}$ defined in (S14), for all $\ell \in \{1, ..., L\}$,*

$$\frac{m_0^{(\ell)} G_0^{(\ell)}(\hat{b}^{(\ell)}; \hat{b}^{(\ell-1)})}{\max \left\{ \sum_{i=1}^{m^{(\ell)}} I(X_i^{(\ell)} > \hat{b}^{(\ell)}), 1 \right\}} = \alpha \{1 + o(1)\}. \tag{S18}$$

*Proof of Proposition 1.* It is suffice to prove that

$$\sup_{i=1}^{m} P(\mathsf{Span}_j(\Omega_i) \notin [\frac{1}{2M_2 m^{1/p}}, \frac{2}{M_1 m^{1/p}}]) \to o(m^{-1}) \; \forall j \in \{1, ..., p\} \tag{S19}$$

Here, $\mathsf{Span}_j(\Omega_i)$ is the length of the sample partition $\Omega_i$ on dimension $j$.

On the $j$th dimension, we conduct the sample quantile partition on all of the $m^{\frac{j-1}{p}}$ sets generated from the previous dimensions. Each of the set contains $N_{(j)} = \frac{N}{m^{(j-1)/p}}$ many cells. Consider we draw $N_{(j)}$ samples $Y_1, ..., Y_{N_{(j)}}$ independently from $UNIF(0, 1/M_1)$, with $m^{1/p}$ and $n_{(j)} = N_j/m^{1/p}$ satisfied the condition C1. Define a sequence of random variables $0 \leq Q_1 \leq ... \leq Q_{m^{1/p}-1} = 1$ as the $1/m^{1/p}, ..., (m^{1/p} - 1)/m^{1/p}$th sample quantile locations. We also define $Q_0 = 0$ and $Q_{m^{1/p}} = 1$.

If exists $C > 0$, s.t. $P(Q_1 > \frac{2n_{(j)}}{M_1 N_{(j)}}) > \frac{C}{m}$, let $S_1 = \sum_{k=1}^{N_{(j)}} I(Y_k \leq \frac{2n_{(j)}}{M_1 N_{(j)}})$, then

$$P(S_1 \leq n_{(j)}) \geq P(Q_1 > \frac{2n_{(j)}}{M_1 N_{(j)}}) > \frac{C}{m}$$

However, based on Chernoff bound, we have

$$P(S_1 \leq n_{(j)}) \leq P\bigl(|S_1 - 2n_{(j)}| > n_{(j)}/3\bigr) = o(m^{-1})$$

Which is a contradiction. Thus, we have

$$P(Q_1 - Q_0 > \frac{2n_{(j)}}{M_1 N_{(j)}}) = o(m^{-1})$$

and similarly,

$$P(Q_{m^{1/p}} - Q_{m^{1/p}-1} > \frac{2n_{(j)}}{M_1 N_{(j)}}) = o(m^{-1})$$

Since for $i = 2, ..., m^{1/p} - 1$,

$$P(Q_i - Q_{i-1} \geq \frac{2n_{(j)}}{M_1 N_{(j)}})$$
$$= \int_0^{\frac{1}{M_1} - \frac{2n_{(j)}}{M_1 N_{(j)}}} P(Q_i - Q_{i-1} \geq \frac{2n_{(j)}}{M_1 N_{(j)}} | Q_{i-1} = y) P(Q_{i-1} = y) dy$$
$$= \int_0^{1 - \frac{2n_{(j)}}{N_{(j)}}} N_{(j)} \binom{N_{(j)} - 1}{(i-1)n_{(j)} - 1} y^{(i-1)n_{(j)}-1} (1-y)^{N_{(j)} - (i-1)n_{(j)}}$$
$$\sum_{s=0}^{n_{(j)}} \binom{N_{(j)} - (i-1)n_{(j)}}{s} \left(\frac{\frac{2n_{(j)}}{N_{(j)}}}{1-y}\right)^s \left(\frac{1 - y - \frac{2n_{(j)}}{N_{(j)}}}{1-y}\right)^{N_{(j)} - (i-1)n_{(j)} - s} dy$$
$$= \sum_{s=0}^{n_{(j)}} \frac{N_{(j)}!}{s!(N_{(j)} - s)!} \left(\frac{2n_{(j)}}{N_{(j)}}\right)^s \left(1 - \frac{2n_{(j)}}{N_{(j)}}\right)^{N_{(j)} - s}$$
$$= P(S_1 \leq n_{(j)}) = o(m^{-1})$$

Thus,

$$m \sup_{i=1}^{m^{1/p}} P((Q_i - Q_{i-1}) \geq \frac{2n_{(j)}}{M_1 N_{(j)}}) = o(1)$$

Let $f(y) = \{N_2 f_2(y) + N_1 f_1(y)\}/(N_2 + N_1)$. It is easy to see that $M_1 \leq f(y) \leq M_2$.

When $N_{(j)}$ samples $Y'_1, ..., Y'_{N_{(j)}}$ are independently drawn from the distribution with density $f : [0, 1] \to [M_1, M_2]$, we have

$$P(Q'_i - Q'_{i-1} \geq \frac{2n_{(j)}}{M_1 N_{(j)}}) = \int_0^{1 - \frac{2n_{(j)}}{M_1 N_{(j)}}} P(Q'_i - Q'_{i-1} \geq \frac{2n_{(j)}}{M_1 N_{(j)}} | Q'_{i-1} = y') P(Q'_{i-1} = y') dy'$$

S8

For any $y_0' \in [0, 1 - \frac{2n_{(j)}}{M_1 N_{(j)}}]$, there always exists corresponding $y_0 = F(y_0')/M_1 \in [0, \frac{1}{M_1} - \frac{2n_{(j)}}{M_1 N_{(j)}}]$, s.t. $M_1 P(Q_{i-1} = y_0) = P(Q_{i-1}' = y_0')/f(y_0')$.

When $y_0' > F^{-1}\left[1 - \frac{2n_{(j)}}{M_1 N_{(j)}}(m^{1/p} - i + 1)\right]$, we have

$$n_{(j)} \leq (m^{1/p} - i + 1)n_{(j)}\beta_{i,U}(y_0') \leq (m^{1/p} - i + 1)n_{(j)}\beta_{i,F}(y_0')$$

Here, $\beta_{i,U}(y_0') = \frac{2n_{(j)}/N_{(j)}}{1 - F(y_0')}$, and $\beta_{i,U}(y_0') = \frac{F(y_0' + 2n_{(j)}/N_{(j)}) - F(y_0')}{1 - F(y_0')}$.

Let,

$$W_{i,F}(y_0') = \sum_{k=1}^{(m^{1/p} - i + 1)n_{(j)}} I(Y_k' \in [y_0', y_0' + \frac{2n_{(j)}}{M_1 N_{(j)}}]) \sim \text{Binom}\left[(m^{1/p} - i + 1)n_{(j)}, \beta_{i,F}(y_0')\right]$$

$$W_{i,U}(y_0') = \sum_{k=1}^{(m^{1/p} - i + 1)n_{(j)}} I(Y_k \in [y_0, y_0 + \frac{2n_{(j)}}{M_1 N_{(j)}}]) \sim \text{Binom}\left[(m^{1/p} - i + 1)n_{(j)}, \beta_{i,U}(y_0)\right]$$

Then

$$P(Q_i' - Q_{i-1}' \geq \frac{2n_{(j)}}{M_1 N_{(j)}} | Q_{i-1}' = y_0') = P(W_{i,F}(y_0') \leq n_{(j)})$$
$$\leq P(W_{i,U}(y_0') \leq n_{(j)})$$
$$= P(Q_i - Q_{i-1} \geq \frac{2n_{(j)}}{M_1 N_{(j)}} | Q_{i-1} = y_0)$$

When $y_0' \leq F^{-1}\left[1 - \frac{2n_{(j)}}{M_1 N_{(j)}}(m^{1/p} - i + 1)\right]$, the corresponding $y_0 = F(y_0')/M_1 \leq \frac{1 - \frac{2n_{(j)}}{M_1 N_{(j)}}(m^{1/p} - i + 1)}{M_1}$, and

$$\mathsf{P}(Q_{i-1}' = y_0') = f(y_{,0})M_1 \mathsf{P}(Q_{i-1} = y_0)$$
$$\leq C\mathsf{P}\left[\sum_{i=1}^{N_{(j)}} I\left(Y_k \geq \frac{1 - \frac{2n_{(j)}}{M_1 N_{(j)}}(m^{1/p} - i + 1)}{M_1}\right) = N_{(j)} - (i-1)n_{(j)}\right]$$
$$\leq C\mathsf{P}\left[\sum_{i=1}^{N_{(j)}} I\left(Y_k \geq \frac{1 - \frac{2n_{(j)}}{M_1 N_{(j)}}(m^{1/p} - i + 1)}{M_1}\right) \leq N_{(j)} - (i-1)n_{(j)}\right]$$
$$= o(m^{-1}) \text{ (Chernoff bound)}$$

S9

Thus,

$$\begin{aligned}
&\mathsf{P}(Q'_i - Q'_{i-1} \geq \frac{2n_{(j)}}{M_1 N_{(j)}}) \\
&= \int_0^{F^{-1}\left[1-\frac{2n_{(j)}}{M_1 N_{(j)}}(m^{1/p}-i+1)\right]} \mathsf{P}(Q'_i - Q'_{i-1} \geq \frac{2n_{(j)}}{M_1 N_{(j)}} | Q'_{i-1} = y')\mathsf{P}(Q'_{i-1} = y')dy' \\
&+ \int_{F^{-1}\left[1-\frac{2n_{(j)}}{M_1 N_{(j)}}(m^{1/p}-i+1)\right]}^{1-\frac{2n_{(j)}}{M_1 N_{(j)}}} \mathsf{P}(Q'_i - Q'_{i-1} \geq \frac{2n_{(j)}}{M_1 N_{(j)}} | Q'_{i-1} = y')\mathsf{P}(Q'_{i-1} = y')dy' \\
&\leq o(m^{-1}) + \mathsf{P}(Q_i - Q_{i-1} \geq \frac{2n_{(j)}}{M_1 N_{(j)}}) \\
&= o(m^{-1})
\end{aligned}$$

Then,

$$\sup_i \mathsf{P}((Q'_i - Q'_{i-1}) \geq \frac{2n_{(j)}}{M_1 N_{(j)}}) = o(m^{-1})$$

Similarly, we can show that

$$\sup_i \mathsf{P}((Q'_i - Q'_{i-1}) \leq \frac{n_{(j)}}{2M_2 N_{(j)}}) = o(m^{-1})$$

□

*Proof of Theorem 1.* For $\breve{X}_i^{(\ell)} \sim \text{Binom}(n_i^{(\ell)}, \theta_i)$, we define $\widetilde{G}_i^{(\ell)}(b; \hat{c}^{(\ell-1)})$ recursively:

$$\widetilde{G}_i^{(\ell)}(b; \hat{c}^{(\ell-1)}) = \mathsf{P}(\breve{X}_i^{(\ell)} > b \mid \widetilde{G}_i^{(\ell)}(X_{i_1}^{(\ell-1)}; \hat{c}^{(\ell-2)}) \geq c^{(\ell-1)}, \widetilde{G}_i^{(\ell)}(\breve{X}_{i_2}^{(\ell-1)}; \hat{c}^{(\ell-2)}) \geq c^{(\ell-1)})$$

With $\widetilde{G}_i^{(1)}(b; \hat{c}^{(0)}) = \mathsf{P}(\breve{X}_i^{(1)} > b)$.

Let $\hat{b}_i^{(\ell)}$ be the value s.t. $G_{0,i}(\hat{b}_i^{(\ell)}; \hat{c}^{(\ell-1)}) = \hat{c}^{(\ell)}$. Then the random variable $FDP^{(\ell)}$ can be decomposed to the product of four parts.

$$FDP^{(\ell)} = \frac{\sum_{i \in \mathcal{B}_{\text{nul}}^{(\ell)}} I(P_{0,i}^{(\ell)} < \hat{c}^{(\ell)})}{\sum_{i \in \mathcal{B}_{\text{nul}}^{(\ell)}} \widetilde{G}_i^{(\ell)}(\hat{b}_i^{(\ell)}; \hat{c}^{(\ell-1)})} \times \frac{\sum_{i \in \mathcal{B}_{\text{nul}}^{(\ell)}} \widetilde{G}_i^{(\ell)}(\hat{b}_i^{(\ell)}; \hat{c}^{(\ell-1)})/m_0^{(\ell)}}{\hat{c}^{(\ell)}}$$

$$\times \frac{\hat{c}^{(\ell)}}{\max(\sum_{1 \leq i \leq m^{(\ell)}} I(P_{0,i}^{(\ell)} < \hat{c}^{(\ell)}), 1)/m^{(\ell)}} \times \frac{m_0^{(\ell)}}{m^{(\ell)}} \quad \text{(S20)}$$

Based on (S20), in order to prove

$$\lim_{N \to +\infty} \mathsf{P}(|FDP^{(\ell)} - \alpha| \leq \epsilon) = 1, \quad \text{(S21)}$$

S10

we only need to prove

$$\mathsf{P}\left\{\left|\frac{m_0^{(\ell)}}{m^{(\ell)}} - 1\right| > \epsilon\right\} \to 0, \quad \text{as } N \to \infty. \tag{S22}$$

$$\mathsf{P}\left\{\left|\frac{\sum_{i \in \mathcal{B}_{\text{nul}}^{(\ell)}} I(P_{0,i}^{(\ell)} < \hat{c}^{(\ell)})}{\sum_{i \in \mathcal{B}_{\text{nul}}^{(\ell)}} \widetilde{G}_i^{(\ell)}(\hat{b}_i^{(\ell)}; \hat{c}^{(\ell-1)})} - 1\right| > \epsilon\right\} \to 0, \quad \text{as } N \to \infty. \tag{S23}$$

$$\mathsf{P}\left\{\left|\frac{\sum_{i \in \mathcal{B}_{\text{nul}}^{(\ell)}} \widetilde{G}_i^{(\ell)}(\hat{b}_i^{(\ell)}; \hat{c}^{(\ell-1)})}{m_0^{(\ell)} \hat{c}^{(\ell)}} - 1\right| > \epsilon\right\} \to 0, \quad \text{as } N \to \infty. \tag{S24}$$

$$\mathsf{P}\left\{\left|\frac{m^{(\ell)} \hat{c}^{(\ell)}}{\max(\sum_{1 \leq i \leq m^{(\ell)}} I(P_{0,i}^{(\ell)} < \hat{c}^{(\ell)}), 1)} - \alpha\right| > \epsilon\right\} \to 0, \quad \text{as } N \to \infty. \tag{S25}$$

We prove (S22) – (S25) by induction.
a) First, let $\ell = 1$.
a1) (S22) holds by Condition C1.
a2) Now we prove (S24).
Given

$$\sum_{i \in \mathcal{B}_{\text{nul}}^{(1)}} \widetilde{G}_i^{(1)}(\hat{b}_i^{(1)}; \hat{c}^{(0)}) - m_0^{(1)} \hat{c}^{(\ell)} =$$

$$\sum_{i \in \mathcal{D}_{\text{nul}}^{(1)}} \left\{\widetilde{G}_i^{(1)}(\hat{b}_i^{(1)}; \hat{c}^{(0)}) - \hat{c}^{(\ell)}\right\} + \sum_{i \in \mathcal{E}_{\text{nul}}^{(1)}} \left\{\widetilde{G}_i^{(1)}(\hat{b}_i^{(1)}; \hat{c}^{(0)}) - \hat{c}^{(\ell)}\right\}, \tag{S26}$$

where $\mathcal{E}_{\text{nul}}^{(1)} = \mathcal{B}_{\text{nul}}^{(1)} \setminus \mathcal{D}_{\text{nul}}^{(1)}$. By Lemma 6 and Lemma 3, we have

$$\sum_{i \in \mathcal{D}_{\text{nul}}^{(1)}} \left\{\widetilde{G}_i^{(1)}(\hat{b}_i^{(1)}; \hat{c}^{(0)})\right\} = m_0^{(1)} \hat{c}^{(\ell)} (1 + o(1)). \tag{S27}$$

By Lemma 8 and Lemma 3,

$$0 \leq \sum_{i \in \mathcal{E}_{\text{nul}}^{(1)}} \left\{\hat{c}^{(\ell)} - \widetilde{G}_i^{(1)}(\hat{b}_i^{(1)}; \hat{c}^{(0)})\right\} \leq \mathsf{Card}(\mathcal{E}_{\text{nul}}^{(1)}) \hat{c}^{(\ell)} = o\{m^{(1)} \hat{c}^{(\ell)}\}. \tag{S28}$$

Combining (S26), (S27), and (S28), we will have (S24).

a3) Based on Lemma 10 and (S27), to prove (S23), it suffices to show that for some $C^{(1)} > 0$,

$$\mathsf{P}\left\{m_0^{(1)} \hat{c}^{(\ell)} \geq C^{(1)} \log m^{(1)}\right\} \to 1. \tag{S29}$$

Recall $\mathcal{G}^{(1)}$ defined in (S4) and $s_G^{(1)} = \mathsf{Card}(\mathcal{G}^{(1)})$. Take a subset $\mathcal{F}^{(1)} \subseteq \mathcal{G}^{(1)}$, such that $s_F^{(1)} = \mathsf{Card}(\mathcal{F}^{(1)}) = r_6 \log m^{(1)}$. For any $i \in \mathcal{F}^{(1)}$, because $\theta_i(1 - \theta_i) \leq 1/4$,

$$n^{(1)} \theta_i - \{n^{(1)} \theta_i (1 - \theta_i)\}^{1/2} \lambda \geq n^{(1)} \theta_0 + \{n^{(1)} \theta_0 (1 - \theta_0)\}^{1/2} \gamma.$$

By (7), $n^{(1)}\theta_0 + \{n^{(1)}\theta_0(1-\theta_0)\}^{1/2}\gamma \geq \hat{b}_i^{(1)}\{1+o(1)\}$ for all $i = 1,...,m$. Combined with Lemma 5, $\tilde{G}_i^{(1)}(\hat{b}_i^{(1)};\hat{c}^{(0)}) > 1 - \{\log m^{(1)}\}^{-1}(\log\log m^{(1)})^{-1/2}$. Then, by Lemma 1,

$$\sum_{i \in \mathcal{F}^{(1)}} P(P_{0,i}^{(1)} \leq \hat{c}^{(1)}) \geq \sum_{i \in \mathcal{F}^{(1)}} \tilde{G}_i^{(1)}(\hat{b}_i^{(1)};\hat{c}^{(0)})(1 - \frac{Cn^2}{N}) > r_6 \log m^{(1)}(1+o(1)). \tag{S30}$$

Let $n^{(1)}\theta_0 - \frac{\{n\theta_0(1-\theta_0)\}^{1/2}}{\sqrt{\log m \log\log m}} = b_{-1} < b_0 < b_1 < ... < b_{\lceil\gamma(\log m \log\log m)^{1/2}\rceil} = n^{(1)}\theta_0 + \{n^{(1)}\theta_0(1-\theta_0)\}^{1/2}\gamma$ satisfy $b_t - b_{t-1} = \frac{\sqrt{n\theta_0(1-\theta_0)}}{\sqrt{\log m \log\log m}}$ for $0 \leq t < \lceil\gamma(\log m \log\log m)^{1/2}\rceil$ and $b_{\lceil\gamma(\log m)^{1/2}\rceil} - b_{\lceil\gamma(\log m \log\log m)^{1/2}\rceil - 1} \leq \frac{\sqrt{n\theta_0(1-\theta_0)}}{\sqrt{\log m \log\log m}}$. For all $i \in \mathcal{F}^{(1)}$, we can get the corresponding p-value $Q_i^{(1)} = \{c_{i,-1}^{(1)}, c_{i0}^{(1)}, ..., c_{i\lceil\gamma(\log m \log\log m)^{1/2}\rceil}^{(1)}\}$, with $c_{ik}^{(1)} = G_{0,i}(b_k;\hat{c}^{(0)})$. After ordering the elements in set $\cup_{i=1}^{|\mathcal{F}^{(1)}|} Q_i^{(1)}$, we get $c_{(0)}^{(1)} < ... < c_{(J)}^{(1)}$, with $J = |\mathcal{F}^{(1)}| \times (\lceil\gamma(\log m \log\log m)^{1/2}\rceil + 1) = o((\log m)^{5.1/2})$. We have:

$$P\Big(\max_{0 \leq j \leq J} \Big|\sum_{i \in \mathcal{F}^{(1)}} I(P_{0,i}^{(1)} < c_{(j)}^{(1)}) - \sum_{i \in \mathcal{F}^{(1)}} P(P_{0,i}^{(1)} < c_{(j)}^{(1)})\Big| \geq 2\sqrt{2}\sqrt{r_6 \log m \log\log m}\Big)$$
$$\leq C(\log m)^{-1/4} \text{ (Azuma's inequality)}$$

For any $c_j^{(1)} \leq c \leq c_{j+1}^{(1)}$, there exist sequences $\delta_3(m) \to 0$, s.t.

$$\Big|\sum_{i \in \mathcal{F}^{(1)}} I(P_{0,i}^{(1)} < c) - \sum_{i \in \mathcal{F}^{(1)}} \mathsf{P}(P_{0,i}^{(1)} < c)\Big|$$
$$\leq \max\Big\{\sum_{i \in \mathcal{F}^{(1)}} \mathsf{P}(P_{0,i}^{(1)} < c_j^{(1)})\Big|\frac{\sum_{i \in \mathcal{F}^{(1)}} I(P_{0,i}^{(1)} < c_j^{(1)})}{\sum_{i \in \mathcal{F}^{(1)}} \mathsf{P}(P_{0,i}^{(1)} < c_j^{(1)})} - 1\Big|(1-\delta_3(m)),$$
$$\sum_{i \in \mathcal{F}^{(1)}} P(P_{0,i}^{(1)} < c_{j+1}^{(1)})\Big|\frac{\sum_{i \in \mathcal{F}^{(1)}} I(P_{0,i}^{(1)} < c_{j+1}^{(1)})}{\sum_{i \in \mathcal{F}^{(1)}} \mathsf{P}(P_{0,i}^{(1)} < c_{j+1}^{(1)})} - 1\Big|(1+\delta_3(m))\Big\} + o(\log m)$$

Thus,

$$\mathsf{P}\Big(\sup_{c \in [\frac{1}{m \log m},\alpha]} \Big|\sum_{i \in \mathcal{F}^{(1)}} I(P_{0,i}^{(1)} < c) - \sum_{i \in \mathcal{F}^{(1)}} \mathsf{P}(P_{0,i}^{(1)} < c)\Big| - o(\log m) \geq 2\sqrt{2r_6 \log m \log\log m}\Big)$$
$$\leq \mathsf{P}\Big(\max_{0 \leq j \leq J} \Big|\sum_{i \in \mathcal{F}^{(1)}} I(P_{0,i}^{(1)} < c_{(j)}^{(1)}) - \sum_{i \in \mathcal{F}^{(1)}} \mathsf{P}(P_{0,i}^{(1)} < c_{(j)}^{(1)})\Big| \geq 2\sqrt{2r_6 \log m \log\log m}\Big)$$
$$= o(1) \tag{S31}$$

Combined with (S30),

$$\mathsf{P}\Big[\sum_{i \in \mathcal{F}^{(1)}} I(P_{0,i}^{(1)} < \hat{c}^{(1)}) \geq r_6 \log m^{(1)}(1+o(1)) - C\{\log m^{(1)} \log\log m^{(1)}\}^{1/2}\Big] \geq 1 - o(1). \tag{S32}$$

S12

It is easy to see that

$$\sum_{1\leq i\leq m^{(1)}} I(P^{(1)}_{0,i} < \hat{c}^{(1)}) \geq \sum_{i\in \mathcal{F}^{(1)}} I(P^{(1)}_{0,i} < \hat{c}^{(1)})$$

Combined with Lemma 11, there exists constant $C^{(1)}$, s.t.

$$\mathsf{P}\left[m_0^{(1)}\hat{c}^{(\ell)} \geq \left\{C^{(1)} \log m^{(1)}\right\}\right] \geq 1 - o(1).$$

a4) Combining Lemma 11 and (S22), (S25) holds.

b) Now we assume (S22) – (S25) hold for layer $1, 2, \ldots, \ell - 1$, and immediately (S21) holds for layer $1, 2, \ldots, \ell - 1$. We will prove (S22) – (S25) hold for layer $\ell$.

b1) First, we prove (S22).

Denote by $V^{(k)}$ and $R^{(k)}$ the numbers of false discoveries and total rejections on layer $k$. Then the total number of true discoveries is $R^{(k)} - V^{(k)}$.

Take any $\tilde{\alpha} > \alpha$, we have

$$\mathsf{P}(\cap_{k=1}^{\ell-1}\{\boldsymbol{x} : FDP^{(k)} < \tilde{\alpha}\}) \to 1. \tag{S33}$$

When $\cap_{k=1}^{\ell-1} FDP^{(k)} \leq \tilde{\alpha}$ holds, $V^{(k)} \leq \tilde{\alpha} R^{(k)}$, and it follows that

$$FD^{(k)} \leq \frac{\tilde{\alpha}}{1+\tilde{\alpha}}(R^{(k)} - V^{(k)}) \leq \frac{\tilde{\alpha}}{1+\tilde{\alpha}} m_1^{(1)}.$$

Then, $m_0^{(\ell)} \geq \frac{m_0^{(1)} - Cm_1^{(1)} - 2^{\ell-1}}{2^{\ell-1}}$. Meanwhile, $m^{(\ell)} \leq \frac{m^{(1)}}{2^{\ell-1}}$. Thus, $m_0^{(\ell)}/m^{(\ell)} \geq 1 - o(1)$ on the space $\cap_{k=1}^{\ell-1}\{\boldsymbol{x} : FDP^{(k)} \leq \tilde{\alpha}\}$. Combined with (S33), we can get (S22).

b2) For (S24), let

$$\mathcal{D}_{\mathrm{nul}}^{(\ell)} = \left\{i \in \{1, \ldots, m_0^{(\ell)}\} : \forall\, j \in S_i^{(\ell)}, \theta_0 - (n \log m^{(1)})^{-1} < \theta_j < \theta_0\right\}.$$

By the proof of (S22) in Part b1), we know that

$$\mathsf{Card}(\mathcal{D}_{\mathrm{nul}}^{(\ell)}) \geq \frac{\mathsf{Card}(\mathcal{D}_{\mathrm{nul}}^{(1)}) - m_1^{(1)} - 2^{\ell-1}}{2^{\ell-1}}.$$

Combined with $\mathsf{Card}(\mathcal{D}_{\mathrm{nul}}^{(1)}) \sim m^{(1)}$, $m_0^{(\ell)} \sim m^{(\ell)} \sim \frac{m^{(1)}}{2^{\ell-1}}$, we have $\mathsf{Card}(\mathcal{D}_{\mathrm{nul}}^{(\ell)}) \sim m_0^{(\ell)} \sim m^{(\ell)}$. Now following the similar argument in a3) we can prove (S24) on layer $\ell$.

b3) Now we prove (S23).

Following the similar argument as in a3), we know that to prove (S23), it suffices to prove that for some constant $C^{(\ell)} > 0$,

$$\mathsf{P}\left\{m_0^{(\ell)}\hat{c}^{(\ell)} \geq C^{(\ell)} \log m^{(1)}\right\} \to 1. \tag{S34}$$

By Condition C3, for any $1 \leq k \leq \ell - 1$, on $\cap_{k=1}^{(\ell-1)} \mathcal{X}^{(k)}$ as defined in (S14),

$$\frac{\hat{b}_i^{(k)} - n^{(k)}\theta_j}{\{n^{(k)}\theta_j(1-\theta_j)\}^{1/2}} \geq (1 - 2^{(k-\ell+1)/2})\beta \geq 0.$$

For any $i' \in \mathcal{G}^{(\ell)}$, for any $1 \leq k \leq \ell$, define, when $N$ is large sufficiently,

$$\mathcal{J}_{i'}^{(k)} = \{i : S_i^{(k)} \subseteq \mathcal{E}_{i'}^{(\ell)}\}.$$

Take $\epsilon > 0$ such that $(1/2 - \epsilon)^{2^L - 1} \geq 2/3$. For any $j \in \mathcal{E}_{i'}^{(\ell)} \in \mathcal{G}^{(\ell)}$,

$$\mathsf{P}\{\mathsf{Card}(\mathcal{J}_{i'}^{(1)}) \geq 2^\ell - 1 \text{ and } \forall i \in \mathcal{J}^{(1)}, X_i^{(1)} \leq \hat{b}_i^{(1)}\} = \mathsf{P}(\max_{j \in \mathcal{E}_{i'}} X_j^{(1)} \leq \hat{b}_i^{(1)}) \geq (1/2 - \epsilon)^{2^\ell - 1}(1 - C\frac{n^2}{N}) \geq 1/3.$$

For any $2 \leq k \leq \ell - 1$, assume

$$\mathsf{P}\{\mathsf{Card}(\mathcal{J}_{i'}^{(k-1)}) \geq 2^{\ell-k+2} - 1 \text{ and } \forall i \in \mathcal{J}^{(k-1)}, X_i^{(k-1)} \leq \hat{b}_i^{(k-1)}\} \geq (1/3)^{k-1}.$$

Then on layer $k$, by Lemma 10 (c),

$$\mathsf{P}\{\mathsf{Card}(\mathcal{J}_{i'}^{(k)}) \geq 2^{\ell-k+1} - 1 \text{ and } \forall i \in \mathcal{J}^{(k)}, X_i^{(k)} \leq \hat{b}_i^{(k)}\}$$
$$\geq \mathsf{P}\{\mathsf{Card}(\mathcal{J}_{i'}^{(k-1)}) \geq 2^{\ell-k+2} - 1 \text{ and } \forall i \in \mathcal{J}_{i'}^{(k-1)}, X_i^{(k-1)} \leq \hat{b}_i^{(k-1)}\}$$
$$\times \mathsf{P}\left[\cap_{i \in \mathcal{J}_{i'}^{(k)}} \{X_i^{(k)} \leq \hat{b}_i^{(k)} \mid X_{i_1}^{(k-1)} \leq \hat{b}_{i_1}^{(k-1)}, X_{i_2}^{(k-1)} \leq \hat{b}_{i_2}^{(k-1)}\}\right]$$
$$\geq \mathsf{P}\{\mathsf{Card}(\mathcal{J}_{i'}^{(k-1)}) \geq 2^{\ell-k+2} - 1 \text{ and } \forall i \in \mathcal{J}_{i'}^{(k-1)}, X_i^{(k-1)} \leq \hat{b}_i^{(k-1)}\}$$
$$\times \mathsf{P}\left[\cap_{i \in \mathcal{J}_{i'}^{(k)}} \{X_i^{(k)} \leq n_i^{(k)}\theta_0 + \sqrt{n_i^{(k)}\theta_0(1-\theta_0)}\beta, X_{i_1}^{(k-1)} \leq n_i^{(k-1)}\theta_0 + \sqrt{n_i^{(k-1)}\theta_0(1-\theta_0)}\beta\right.$$
$$\left., X_{i_2}^{(k-1)} \leq n_i^{(k-1)}\theta_0 + \sqrt{n_i^{(k-1)}\theta_0(1-\theta_0)}\beta\}\right]$$
$$\geq (1/3)^{k-1} \times (1/3) = (1/3)^k.$$

Therefore,

$$\mathsf{P}\{\mathsf{Card}(\mathcal{J}_{i'}^{(\ell-1)}) \geq 3 \text{ and } \forall i \in \mathcal{J}_{i'}^{(\ell-1)}, X_i^{(\ell-1)} \leq \hat{b}_i^{(\ell-1)}\} \geq (1/3)^{\ell-1}.$$

Let $\mathcal{F}^{(\ell)}$ be the subset of $\mathcal{G}^{(\ell)}$ such that $s_F^{(\ell)} = \mathsf{Card}(\mathcal{F}^{(\ell)}) = r_6 \log m^{(1)}/2^L$, and elements in $\mathcal{F}^{(\ell)}$ are mutually disjoint. Also let

$$\widetilde{\mathcal{F}}^{(\ell)} = \left\{i : i \in \mathcal{F}^{(\ell)} \text{ and there exists some } i' \text{ such that } S_{i'}^{(\ell)} \subseteq \mathcal{E}_i^{(\ell)}\right\}.$$

By Chebyshev's inequality

$$\mathsf{P}\left\{\left|\mathsf{Card}(\widetilde{\mathcal{F}}^{(\ell)}) - \sum_{i \in \mathcal{F}^{(\ell)}} \mathsf{P}(i \in \widetilde{\mathcal{F}}^{(\ell)})\right| > \left\{\sum_{i \in \mathcal{F}^{(\ell)}} \mathsf{P}(i \in \widetilde{\mathcal{F}}^{(\ell)})\right\}^{1/2} (\log m^{(1)})^{1/4}\right\}$$
$$\leq \frac{\sum_{i \in \mathcal{F}^{(\ell)}} \mathsf{P}(i \in \widetilde{\mathcal{F}}^{(\ell)})\{1 - \mathsf{P}(i \in \widetilde{\mathcal{F}}^{(\ell)})\}}{(\log m^{(1)})^{1/2} \sum_{i \in \mathcal{F}^{(\ell)}} \mathsf{P}(i \in \widetilde{\mathcal{F}}^{(\ell)})} \leq (\log m^{(1)})^{-1/2}.$$

Define
$$\mathcal{X}_2 = \cup_{h=1}^{\ell-1} \mathcal{X}^{(h)} \cap \{\mathsf{Card}(\widetilde{\mathcal{F}}^{(\ell)})\} \geq c(\log m^{(1)})\}. \tag{S35}$$

Then $\mathsf{P}(\mathcal{X}_2) \geq 1 - o(1)$.

On $\mathcal{X}_2$, By Condition C3,
$$n^{(\ell)}\theta_j - \{n^{(\ell)}\theta_j(1-\theta_j)\}^{1/2}\lambda > n^{(\ell)}\theta_0 + \{n^{(\ell)}\theta_0(1-\theta_0)\}^{1/2}\gamma \geq \hat{b}_j^{(\ell)}(1+o(1)).$$

By Lemma 5 and Lemma 10,
$$\sum_{i \in \mathcal{F}^{(\ell)}} \widetilde{G}_i^{(\ell)}(\hat{b}_i^{(\ell)}; \hat{c}^{(\ell-1)}) \geq \mathsf{Card}(\widetilde{\mathcal{F}}^{(\ell)})(1+o(1)) \geq c\log m^{(1)}.$$

Following similar argument from (S32) in a2), we can get (S34).

b4) Finally, by combining Lemma 11 and (S22), we have (S25).

□

*Proof of Theorem 2.* $\forall i$, let $j_i^{(\ell)} = \arg\sup_{j \in S_i} \theta_j$. If $i \in \mathcal{W}^{(\ell)}$, then $\theta_{j_i^{(\ell)}} > \theta_0$. Let

$$\mathcal{X}_3 = \left\{\boldsymbol{x} : \inf_{\ell \in \{1,\ldots,L\}} \inf_{i \in \mathcal{W}^{(\ell)}} \theta_{j_i^{(\ell)}} \geq \theta_0 + C_i^{(\ell)}(n^{(1)}\log m)^{-1}\right\} \tag{S36}$$

If for some $C_i^{(\ell)}$ (may depending on $i$ and $\ell$), $\mathsf{P}(\mathcal{X}_3) \to 1$, then by Condition C4, we know that
$$\mathsf{P}(\forall \ell = 1, \ldots, L \text{ and } i \in \mathcal{W}^{(\ell)}, \text{we have } S_i^{(\ell)} \subseteq \mathcal{H}_{\text{alt}}) \to 1.$$

Subsequently,
$$\mathsf{P}\big\{\forall \ell = 1, \ldots, L, \ \mathsf{Card}(\mathcal{R}_{\text{nul}}^{(\ell)}) = 2^{\ell-1}V^{(\ell)}\big\} \to 1.$$

Combined with Theorem 1, we can prove the overall $FDP^{1:L}$ is constant to $\alpha$.

Now it suffices to prove that for some $C_i^{(\ell)}$, $\mathcal{P}(\mathcal{X}_3) \to 1$.

For any $i$ with $S_i^{(\ell)} \cap \mathcal{H}_{\text{alt}} \neq \emptyset$, consider the event $\theta_{j_i^{(\ell)}} - \theta_0 \leq (n^{(1)}\log m)^{-1}\varepsilon_N$, with some $\varepsilon_N \to 0$. $\forall j \in S_i^{(\ell)}$, $\theta_j \leq \theta_{j_i^{(\ell)}}$. Then by the similar arguments as Lemma 8, Lemma 10, and Lemma 5,

$$\mathsf{P}(i \in \mathcal{W}^{(\ell)} \mid S_i^{(\ell)} \cap \mathcal{H}_{\text{alt}} \neq \emptyset, \theta_{j_i^{(\ell)}} - \theta_0 \leq (n^{(1)}\log m)^{-1}\varepsilon_N, \mathcal{X})$$
$$= \mathsf{P}(X_i^{(\ell)} > \hat{b}_i^{(\ell)} \mid S_i^{(\ell)} \cap \mathcal{H}_{\text{alt}} \neq \emptyset, \theta_{j_i^{(\ell)}} - \theta_0 \leq (n^{(1)}\log m)^{-1}\varepsilon_N, \mathcal{X})$$
$$\leq \mathsf{P}(X_i^{(\ell)} > \hat{b}_i^{(\ell)} \mid \forall j \in S_i^{(\ell)}, \theta_j = \theta_{j_i^{(1)}} \leq (n^{(1)}\log m)^{-1}\varepsilon_N, \mathcal{X})$$
$$\leq o(1).$$

S15

Then
$$\mathsf{P}(i \in \mathcal{W}^{(\ell)} = \mathcal{H}_{\text{alt}} \cap \mathcal{U}^{(\ell)}, \theta_{j_i^{(\ell)}} - \theta_0 \leq (n^{(1)} \log m)^{-1} \varepsilon_N, \mathcal{X}) \leq o(1).$$

Then
$$\mathsf{P}(\mathcal{X}_3) \geq 1 - \mathsf{Card}(\mathcal{H}_{\text{alt}})(m^{(1)})^{-(1-r_1)} - P(\mathcal{X}^c) \geq 1 - o(1).$$

□

## S4. Proof of Lemmas

*Proof of Lemma 1.* Consider a partition on the sample space $\Omega$ with $m$ bins $\{\Omega_1, \ldots, \Omega_m\}$. On such a partition, $N_1$ samples from cohort 1 and $N_2$ samples from cohort 2 are collected. The total number of samples falling into bin $i$ is $n_i$, $i = 1, \ldots, m$, with $\tilde{X}_i$ from cohort 1 and $X_i$ from cohort 2. Without loss of generality, let's consider the joint distribution of $(X_1, \ldots, X_K)$ given $n_1, \ldots, n_K$ and $N_1, N_2$.

By Bayes's equality,

$$\begin{aligned}
&\mathsf{P}(X_1 = x_1', \ldots, X_K = x_K' \mid n_1, \ldots, n_K, N_1, N_2) \\
&= \frac{\mathsf{P}(X_1 = x_1', \ldots, X_K = x_K', n_1, \ldots, n_K \mid N_1, N_2)}{\mathsf{P}(n_1, \ldots, n_K \mid N_1, N_2)} \\
&= \frac{\mathsf{P}(\tilde{X}_1 = \tilde{x}_1', \ldots, \tilde{X}_K = \tilde{x}_K', X_1 = x_1', \ldots, X_K = x_K', \mid N_1, N_2)}{\sum_{\substack{x_1 + \tilde{x}_1 = n_1 \\ \vdots \\ x_K + \tilde{x}_K = n_K}} \mathsf{P}(X_1 = x_1, \ldots, X_K = x_K, \tilde{X}_1 = x_1, \ldots, \tilde{X}_K = \tilde{x}_K \mid N_1, N_2)}
\end{aligned} \quad (S37)$$

It is equivalent to consider that $N_1/N_2$ samples have been partitioned into $K+1$ bins with bin $i$ containing $\tilde{X}_i/X_i$ samples such that $\tilde{X}_i + X_i = n_i$, and $\sum_{i=1}^{K+1} \tilde{X}_i = N_1$ and $\sum_{i=1}^{K} X_i = N_2$. For bin $i$ with $i \in \{1, \ldots, K\}$, let $\tilde{p}_i = \int_{\Omega_i} f_1(y) dy$ and $p_i = \int_{\Omega_i} f_2(y) dy$, where $f_1$ and $f_2$ are PDF of the two cohorts. Also, let $\tilde{p}_{K+1} = 1 - \sum_{i=1}^{K} \int_{\Omega_i} f_1(y) dy$, and $p_{K+1} = 1 - \sum_{i=1}^{K} \int_{\Omega_i} f_2(y) dy$. Further, let

$$q_i = \frac{N_1}{N} \tilde{p}_i + \frac{N_2}{N} p_i, \quad \tilde{\theta}_i = \frac{\frac{N_1}{N} \tilde{p}_i}{q_i}, \quad \theta_i = \frac{\frac{N_2}{N} p_i}{q_i}$$

It is easy to see that when $M_1 \leq \inf_{s \in \{1,2\}} \inf_y f_s(y) \leq \sup_{s \in \{1,2\}} \sup_y f_s(y) \leq M_2$, $\sum_{i=1}^{K} \leq CKn^{(1)}/N$ and $\sum_{i=1}^{K} \tilde{p}_i \leq CKn^{(1)}/N$.

The definition of $\theta_1, \ldots, \theta_K$ are the same as in (1). Also we have $\theta_i + \tilde{\theta}_i = 1$, for

$i = 1, \ldots, K+1$.

$$\mathsf{P}(X_1 = x_1, \ldots, X_K = x_K, \tilde{X}_1 = x_1, \ldots, \tilde{X}_K = x_K \mid N_1, N_2)$$
$$= \frac{N_1!}{\prod_{i=1}^{K+1} \tilde{x}_i!} \prod_{i=1}^{K+1} \tilde{p}_i^{\tilde{x}_i} \cdot \frac{N_2!}{\prod_{i=1}^{K+1} x_i!} \prod_{i=1}^{K+1} p_i^{x_i}$$
$$= \frac{N_1!}{\prod_{i=1}^{K+1} \tilde{x}_i!} \prod_{i=1}^{K+1} \left(\frac{N_1}{N} \tilde{p}_i\right)^{\tilde{x}_i} \cdot \frac{N_2!}{\prod_{i=1}^{K+1} x_i!} \prod_{i=1}^{K+1} \left(\frac{N_2}{N} p_i\right)^{x_i} \cdot \left(\frac{N}{N_1}\right)^{N_1} \left(\frac{N}{N_2}\right)^{N_2}$$
$$= \frac{N_1!}{\prod_{i=1}^{K+1} \tilde{x}_i!} \prod_{i=1}^{K+1} \tilde{\theta}_i^{\tilde{x}_i} \cdot \frac{N_2!}{\prod_{i=1}^{K+1} x_i!} \prod_{i=1}^{K+1} \theta_i^{x_i} \cdot \left(\frac{N}{N_1}\right)^{N_1} \left(\frac{N}{N_2}\right)^{N_2} \cdot \prod_{i=1}^{K+1} q_i^{n_i}$$
$$= \left(\frac{N}{N_1}\right)^{N_1} \left(\frac{N}{N_2}\right)^{N_2} N_1! N_2! \prod_{i=1}^{K+1} q_i^{n_i} \cdot \frac{1}{\prod_{i=1}^{K} \tilde{x}_i!} \prod_{i=1}^{K} \tilde{\theta}_i^{\tilde{x}_i} \cdot \frac{1}{\prod_{i=1}^{K} x_i!} \prod_{i=1}^{K} \theta_i^{x_i} \cdot \frac{\tilde{\theta}_{K+1}^{\tilde{x}_{K+1}} \theta_{K+1}^{x_{K+1}}}{\tilde{x}_{K+1}! x_{K+1}!}$$

Now let

$$g(\tilde{x}_1, \ldots, \tilde{x}_K, x_1, \ldots, x_K) = \frac{1}{\prod_{i=1}^{K} \tilde{x}_i!} \prod_{i=1}^{K} \tilde{\theta}_i^{\tilde{x}_i} \cdot \frac{1}{\prod_{i=1}^{K} x_i!} \prod_{i=1}^{K} \theta_i^{x_i},$$

$$w\{(\tilde{x}_i, x_i, \tilde{x}'_i, x'_i)_{i=1\ldots,m}\} = \frac{\tilde{x}'_{K+1}! x'_{K+1}!}{\tilde{x}_{K+1}! x_{K+1}!} \frac{\tilde{\theta}_{K+1}^{\tilde{x}_{K+1}} \theta_{K+1}^{x_{K+1}}}{\tilde{\theta}_{K+1}^{\tilde{x}'_{K+1}} \theta_{K+1}^{x'_{K+1}}}.$$

Then,

$$(\text{S37}) = \frac{g(\tilde{x}'_1, \ldots, \tilde{x}'_K, x'_1, \ldots, x'_K)}{\sum_{\substack{x_1+\tilde{x}_1=n_1 \\ \vdots \\ x_K+\tilde{x}_K=n_K}} g(\tilde{x}_1, \ldots, \tilde{x}_K, x_1, \ldots, x_K) w(\tilde{x}_1, \ldots, \tilde{x}_K, x_1, \ldots, x_K)}$$

We now show that the term $w\{(\tilde{x}_i, x_i, \tilde{x}'_i, x'_i)_{i=1\ldots,m}\}$ is close to 1. Let

$$\log w\{(\tilde{x}_i, x_i, \tilde{x}'_i, x'_i)_{i=1\ldots,m}\} = A\{(\tilde{x}_i, x_i, \tilde{x}'_i, x'_i)_{i=1\ldots,m}\} + B\{(\tilde{x}_i, x_i, \tilde{x}'_i, x'_i)_{i=1\ldots,m}\},$$

where

$$A\{(\tilde{x}_i, x_i, \tilde{x}'_i, x'_i)_{i=1\ldots,m}\} = \log\left(\frac{\tilde{x}'_{K+1}!}{\tilde{x}_{K+1}!}\right) + \log\left(\frac{x'_{K+1}!}{x_{K+1}!}\right)$$
$$B\{(\tilde{x}_i, x_i, \tilde{x}'_i, x'_i)_{i=1\ldots,m}\} = (\tilde{x}_{K+1} - \tilde{x}'_{K+1}) \log \tilde{\theta}_{K+1} + (x_{K+1} - x'_{K+1}) \log \theta_{K+1}$$

Let $\delta_{K+1} = \tilde{x}_{K+1} - \tilde{x}'_{K+1}$. Because $n_{K+1} = \tilde{x}'_{K+1} + x'_{K+1} = \tilde{x}_{K+1} + x_{K+1}$, we have $x'_{K+1} - x_{K+1} = \delta_{K+1}$. It is also easy to see that $|\delta_{K+1}| \leq Kn^{(1)}$.

When $\delta_{K+1} = 0$, $w\{(\tilde{x}_i, x_i, \tilde{x}'_i, x'_i)_{i=1\ldots,m}\} = 1$.

Now assume $\delta_{K+1} > 0$. Then

$$A\{(\tilde{x}_i, x_i, \tilde{x}'_i, x'_i)_{i=1\ldots,m}\}$$
$$= -\sum_{i=1}^{\delta_{K+1}} \log(\tilde{x}'_{K+1} + i) + \sum_{i=1}^{\delta_{K+1}} \log(x_{K+1} + i)$$
$$= \sum_{i=1}^{\delta_{K+1}} \log\left(\frac{x_{K+1} + i}{\tilde{x}'_{K+1} + i}\right) = \sum_{i=1}^{\delta_{K+1}} \log\left(\frac{N_2 - \sum_{i=1}^{K} x_i + i}{N_1 - \sum_{i=1}^{K} \tilde{x}'_i + i}\right)$$
$$= \delta_{K+1} \log\left(\frac{N_2}{N_1}\right) + \sum_{i=1}^{\delta_{K+1}} \left\{\log\left(1 - \frac{\sum_{i=1}^{K} x_i + i}{N_2}\right) + \log\left(1 + \frac{\sum_{i=1}^{K} \tilde{x}'_i - i}{N_1 - \sum_{i=1}^{K} \tilde{x}'_i + i}\right)\right\}.$$

By Taylor expansions, we know that $|\log(1 + x) - x| \leq Cx^2$ and $|\log(1 - x) + x| \leq Cx^2$. Plugging the expansion bound into the above expression, we have

$$A\{(\tilde{x}_i, x_i, \tilde{x}'_i, x'_i)_{i=1\ldots,m}\} = \delta_{K+1} \log\left(\frac{N_2}{N_1}\right) \pm CK(n^{(1)})^2/N. \tag{S38}$$

When $\delta_{K+1} < 0$, We can show (S38) similarly.

On the other hand,

$$B\{(\tilde{x}_i, x_i, \tilde{x}'_i, x'_i)_{i=1\ldots,m}\} = \delta_{K+1} \log\left(\frac{\tilde{\theta}_{K+1}}{\theta_{K+1}}\right)$$
$$= \delta_{K+1} \log\left(\frac{N_1}{N_2}\right) + \delta_{K+1} \log\left(\frac{1 - \sum_{i=1}^{K} \tilde{p}_i}{1 - \sum_{i=1}^{K} p_i}\right)$$
$$= \delta_{K+1} \log\left(\frac{N_1}{N_2}\right) \pm CK^2(n^{(1)})^2/N.$$

Combining the bounding results for $A\{(\tilde{x}_i, x_i, \tilde{x}'_i, x'_i)_{i=1\ldots,m}\}$ and $B\{(\tilde{x}_i, x_i, \tilde{x}'_i, x'_i)_{i=1\ldots,m}\}$, we have

$$|w\{(\tilde{x}_i, x_i, \tilde{x}'_i, x'_i)_{i=1\ldots,m}\}| \simeq \exp\{C(n^{(1)})^2/N\} \simeq 1 + CK^2(n^{(1)})^2/N.$$

This leads to

$$\text{(S37)} \simeq \frac{g(\tilde{x}'_1, \ldots, \tilde{x}'_K, x'_1, \ldots, x'_K)}{\sum_{\substack{x_1 + \tilde{x}_1 = n_1 \\ \vdots \\ x_K + \tilde{x}_K = n_K}} g(\tilde{x}_1, \ldots, \tilde{x}_K, x_1, \ldots, x_K)} \cdot \left\{1 + CK^2(n^{(1)})^2/N\right\}.$$

Then,

$$\sum_{\substack{x_1 + \tilde{x}_1 = n_1 \\ \vdots \\ x_K + \tilde{x}_K = n_K}} g(\tilde{x}_1, \ldots, \tilde{x}_K, x_1, \ldots, x_K) \simeq \prod_{i=1}^{K} \sum_{\tilde{x}_i + x_i = n_i} \frac{\tilde{\theta}_i^{\tilde{x}_i} \theta_i^{x_i}}{x_i! \tilde{x}_i!} = \prod_{i=1}^{K} \frac{1}{n_i!}.$$

Therefore

$$\text{(S37)} \simeq \prod_{i=1}^{K} \frac{n_i!}{x_i! \tilde{x}_i!} \tilde{\theta}_i^{\tilde{x}_i} \theta_i^{x_i} \cdot \left\{1 + C(n^{(1)})^2/N\right\}.$$

$\square$

S18

*Proof of Lemma 2.* Note that $\theta_i = \frac{N_2 \int_{\Omega_i} f_2(y)\mathrm{d}y}{N_2 \int_{\Omega_i} f_2(y)\mathrm{d}y + N_1 \int_{\Omega_i} f_1(y)\mathrm{d}y}$. To prove (S2), it suffices to show that

$$\left| \frac{\int_{\Omega_j} f_2(y)\mathrm{d}y}{\int_{\Omega_j} f_1(y)\mathrm{d}y} - \frac{\int_{\Omega_{j-1}} f_2(y)\mathrm{d}y}{\int_{\Omega_{j-1}} f_1(y)\mathrm{d}y} \right| \leq C/m.$$

Let $y_{s,i,\max} = \arg\max_{\Omega_i} f_s(y)$ and $y_{s,i,\min} = \arg\min_{\Omega_i} f_s(y)$, for $s \in \{1,2\}$. Then

$$\left| \frac{\int_{\Omega_i} f_2(y)\mathrm{d}y}{\int_{\Omega_i} f_1(y)\mathrm{d}y} - \frac{\int_{\Omega_{i-1}} f_2(y)\mathrm{d}y}{\int_{\Omega_{i-1}} f_1(y)\mathrm{d}y} \right|$$
$$\leq \left| \frac{f_2(y_{2,i,\max})}{f_1(y_{1,i,\min})} - \frac{f_2(y_{2,i-1,\min})}{f_1(y_{1,i-1,\max})} \right|$$
$$\leq \frac{|f_2(y_{2,i,\max}) - f_2(y_{2,i-1,\min})| f_1(y_{1,i-1,\max}) + |f_1(y_{1,i-1,\max}) - f_1(y_{1,i,\min})| f_2(y_{2,i-1,\min})}{f_1(y_{1,j,\min}) f_1(y_{1,j-1,\max})}$$
$$\leq \frac{M_2}{M_1^2} \max\{|f_2(y_{2,i,\max}) - f_2(y_{2,i-1,\min})|, |f_1(y_{1,i-1,\max}) - f_1(y_{1,i,\min})|\}. \tag{S39}$$

Based on the multivariate Taylor expansion,

$$|f_2(y_{2,i,\max}) - f_2(y_{2,i-1,\min})| \leq C \sup_{y \in \Omega} \|\nabla f_2(y)\|_2 \|y_{2,i,\max} - y_{2,i-1,\min}\|_2$$
$$\leq CM_3(\mathsf{Span}(\Omega_{i-1}) + \mathsf{Span}(\Omega_i))^p \leq C/m.$$

Similarly, we can show that

$$|f_1(y_{1,i-1,\max}) - f_1(y_{1,i,\min})| \leq C/m.$$

Combined with (S39), we get the conclusion (S2). $\square$

*Proof of Lemma 3.* Based on the partition process, for any bin $\Omega_i$,

$$\sum_{k=1}^{N} I(\text{Cell } k \text{ falls into } \Omega_i) = n.$$

Taking expectation on both sides, we have

$$N \cdot \mathsf{P}(\text{Cell } k \text{ falls into } \Omega_i) = \int_{\Omega_i} \{N_2 f_2(y) + N_1 f_1(y)\} \mathrm{d}y = n.$$

Combined with the definition of $\theta_i$ in (1), we have

$$n^{(1)} \sum_{i=1}^{m^{(1)}} \theta_i = N_2 \sum_{i=1}^{n^{(1)}} \int_{\Omega_i} f_2(y)\mathrm{d}y = N_2 = N\theta_0.$$

It leads to
$$\sum_{i=1}^{m^{(1)}} \theta_i = \sum_{i \in \mathcal{H}_{\text{nul}}^{(1)}} \theta_i + \sum_{i \in \mathcal{H}_{\text{alt}}^{(1)}} \theta_i = m^{(1)} \theta_0.$$

Now let $\delta = \sup_{i \in \mathcal{H}_{\text{alt}}^{(1)}} (\theta_i - \theta_0)$. Then
$$m_0^{(1)} \theta_0 - \sum_{i \in \mathcal{H}_{\text{nul}}^{(1)}} \theta_i = \sum_{i \in \mathcal{H}_{\text{alt}}^{(1)}} (\theta_i - \theta_0) \leq m_1^{(1)} \delta.$$

Because $\theta_i \leq \theta_0$ in $\mathcal{H}_{\text{nul}}^{(1)}$ and $\mathcal{E}_{\text{nul}}^{(1)} = \mathcal{H}_{\text{nul}}^{(1)} \setminus \mathcal{D}_{\text{nul}}^{(1)} \subseteq \mathcal{H}_{\text{nul}}^{(1)}$,
$$\sum_{i \in \mathcal{E}_{\text{nul}}^{(1)}} (\theta_i - \theta_0) \leq m_1^{(1)} \delta.$$

On $\mathcal{E}_{\text{nul}}^{(1)}$, $\theta_i - \theta_0 \geq (n^{(1)} \log m^{(1)})^{-1}$. Combined with Condition C1,
$$\text{Card}(\mathcal{E}_{\text{nul}}^{(1)}) \leq m_1^{(1)} \delta (n^{(1)} \log m^{(1)}) \leq r_2 \delta \{m^{(1)}\}^{\frac{r_4}{1-r_4} + r_1} \log m^{(1)} = o(m^{(1)}).$$

Thus,
$$\frac{\text{Card}(\mathcal{D}_{\text{nul}}^{(1)})}{m_0^{(1)}} = 1 - \frac{\text{Card}(\mathcal{E}_{\text{nul}}^{(1)})}{m_0^{(1)}} = 1 - o(1).$$

□

*Proof of Lemma 4 b).* Given $|n' - n|$ is uniformly bounded by $n\delta_0(m, n)$, $|\frac{n}{n'} - 1|$ is uniformly bounded by $\delta_0'(m, n) = o\left(\sqrt{\frac{1}{n \log m}}\right)$.

Based on part a), we have
$$\mathsf{P}(\check{X}' = k) = \frac{1}{\sqrt{2\pi n' \theta (1-\theta)}} \exp\left(-\frac{(\check{x}' - n'\theta)^2}{2n'\theta(1-\theta)}\right) \cdot (1 + \epsilon_{n'}(k))$$

Thus, there exists $\epsilon_{n,1}' = o(1)$, s.t.
$$\frac{\mathsf{P}(\check{X}' = k)}{\sqrt{2\pi n \theta (1-\theta)}} \exp\left(-\frac{(\check{x}' - n\theta)^2}{2n\theta(1-\theta)}\right)$$
$$\leq \delta_0(m,n) \exp\left\{\delta_0(m,n) \frac{(\check{X}' - n\theta)^2}{n\theta(1-\theta)} + n\delta_0^2(m,n) + 2\sqrt{n}\delta_0(m,n)\sqrt{1 + \delta_0'(m,n)} \frac{|\check{X}' - n\theta|}{\sqrt{n\theta(1-\theta)}}\right\}$$
$$\leq 1 + \epsilon_{n,1}'$$

Similarly, by some calculation, there exists $\epsilon_{n,2}' = o(1)$ with
$$\frac{\mathsf{P}(\check{X}' = k)}{\sqrt{2\pi n \theta (1-\theta)}} \exp\left(-\frac{(\check{x}' - n\theta)^2}{2n\theta(1-\theta)}\right) \geq 1 - \epsilon_{n,2}'$$

□

*Proof of Lemma 6.* Let $\breve{X}_{S_i^{(\ell)}} = \{\breve{X}_j : j \in S_i^{(\ell)}\}$, with $\breve{X}_j \sim \text{Binom}(n_j, \theta_j)$

For any sample space $\Omega$, we have

$$\mathsf{P}(\breve{X}_{S_i^{(\ell)}} \in \Omega \mid S_i^{(\ell)} \subseteq \mathcal{D}_{\text{nul}}) - \mathsf{P}(\breve{X}_{S_i^{(\ell)}} \in \Omega \mid \theta_j = \theta_0, \ \forall \ j \in S_i^{(\ell)})$$

$$= \sum_{\breve{x}_{S_i^{(\ell)}} \in \Omega} \binom{n_j}{\breve{x}_j} \theta_j^{\breve{x}_j}(1-\theta_j)^{n_j-\breve{x}_j} - \sum_{\breve{x}_{S_i^{(\ell)}} \in \Omega} \binom{n_j}{\breve{x}_j} \theta_0^{\breve{x}_j}(1-\theta_0)^{n_j-\breve{x}_j}$$

$$= \sum_{\breve{x}_{S_i^{(\ell)}} \in \Omega} \binom{n_j}{\breve{x}_j} \theta_0^{\breve{x}_j}(1-\theta_0)^{n_j-\breve{x}_j} A_j, \tag{S40}$$

where $A_j = \left(\frac{\theta_j}{\theta_0}\right)^{\breve{x}_j} \left(\frac{1-\theta_j}{1-\theta_0}\right)^{n_j-\breve{x}_j} - 1$.

Let $\theta_j = \theta_0 - a_j$. If $j \in \mathcal{D}_{\text{nul}}^{(1)}$, $0 \le a_j < (n^{(1)} \log m^{(1)})^{-1}$. Thus

$$\max_{j \in \mathcal{D}_{\text{nul}}^{(1)}} |A_j| \le \max_{j \in \mathcal{D}_{\text{nul}}^{(1)}} \left| \left(1 - \frac{a_j}{\theta_0}\right)^{\breve{x}_j} \left(1 + \frac{a_j}{1-\theta_0}\right)^{n_j-\breve{x}_j} - 1 \right| \le \max_{j \in \mathcal{D}_{\text{nul}}^{(1)}} \left| \left(1 + \frac{a_j}{1-\theta_0}\right)^{n_j-\breve{x}_j} - 1 \right|$$

$$\le 2n \max_{j \in \mathcal{H}^{(\ell)}} \left| \frac{a_j}{1-\theta_0} \right| \le Cn(n^{(1)} \log m^{(1)})^{-1} = C(\log m^{(1)})^{-1}.$$

Combining with (S40), we have

$$\max_{S_i^{(\ell)} \subseteq \mathcal{D}_{\text{nul}}} \frac{\left| \mathsf{P}(\breve{X}_{S_i^{(\ell)}} \in \Omega_i^{(\ell)} \mid S_i^{(\ell)} \subseteq \mathcal{D}_{\text{nul}}) - \mathsf{P}(\breve{X}_{S_i^{(\ell)}} \in \Omega \mid \theta_j = \theta_0, \ \forall \ j \in S_i^{(\ell)}) \right|}{\mathsf{P}(\breve{X}_{S_i^{(\ell)}} \in \Omega \mid \theta_j = \theta_0, \ \forall \ j \in S_i^{(\ell)})} \le C(\log m^{(1)})^{-1}. \tag{S41}$$

Let

$$\widetilde{PA}_i = \mathsf{P}(X_i^{(\ell)} > b_i^{(\ell)}, \breve{X}_{i1}^{(\ell-1)} \le b_{i1}^{(\ell-1)}, \breve{X}_{i2}^{(\ell-1)} \le b_{i2}^{(\ell-1)} \mid S_i^{(\ell)} \subseteq \mathcal{D}_{\text{nul}}^{(1)})$$

$$PA_i = \mathsf{P}(\breve{X}_i^{(\ell)} > b_i^{(\ell)}, \breve{X}_{i1}^{(\ell-1)} \le b_{i1}^{(\ell-1)}, \breve{X}_{i2}^{(\ell-1)} \le b_{i2}^{(\ell-1)} \mid \theta_j = \theta_0, \forall \ j \in S_i^{(\ell)})$$

$$\widetilde{PB}_i = \mathsf{P}(\breve{X}_{i1}^{(\ell-1)} \le b_{i1}^{(\ell-1)}, \breve{X}_{i2}^{(\ell-1)} \le b_{i2}^{(\ell-1)} \mid S_i^{(\ell)} \subseteq \mathcal{D}_{\text{nul}}^{(1)}),$$

$$PB_i = \mathsf{P}(\breve{X}_{i1}^{(\ell-1)} \le b_{i1}^{(\ell-1)}, \breve{X}_{i2}^{(\ell-1)} \le b_{i2}^{(\ell-1)} \mid \theta_j = \theta_0, \forall j \in S_i^{(\ell)}).$$

By (S41), we have
$$\widetilde{PA}_i = PA_i(1 + a_i), \quad \widetilde{PB}_i = PB_i(1 + b_i),$$

with

$$\max_{S_i^{(\ell)} \subseteq \mathcal{D}_{\text{nul}}^{(1)}} \{\max(|a_i|, |b_i|)\} \le C(\log m^{(1)})^{-1}. \tag{S42}$$

Thus,

$$\left| \tilde{G}_i^{(\ell)}(b_i^{(\ell)}; c^{(\ell-1)}) - G_{0,i}^{(\ell)}(b_i^{(\ell)}; c^{(\ell-1)}) \right| = \left| \frac{\widetilde{PA}_i}{\widetilde{PB}_i} - \frac{PA_i}{PB_i} \right|$$

$$= \left| \frac{PA_i(1+a_i)}{PB_i(1+b_i)} - \frac{PA_i}{PB_i} \right| = G_{0,i}^{(\ell)}(b_i^{(\ell)}; c^{(\ell-1)}) \left| \frac{a_i - b_i}{1 + b_i} \right|.$$

S21

Together with (S42), we have

$$\max_{S_i^{(\ell)} \subseteq \mathcal{D}_{\mathrm{nul}}} \frac{\left|\tilde{G}_i^{(\ell)}(b_i^{(\ell)}; c^{(\ell-1)}) - G_{0,i}^{(\ell)}(b_i^{(\ell)}; c^{(\ell-1)})\right|}{G_{0,i}^{(\ell)}(b_i^{(\ell)}; c^{(\ell-1)})} \leq C(\log m^{(1)})^{-1}$$

$\square$

*Proof of Lemma 7.* Define a set $\mathcal{D}_b = \{b : |b - n_i\theta_0| < 2\sqrt{n_i \log m \theta_0(1-\theta_0)}\}$.

Let $Z \sim N(0,1)$. Based on Lemma 4 and Lemma 5, we have,

$$\sup_{b \in \mathcal{D}_b} \left| \frac{G_{0,i}^{(1)}(b; \hat{c}^{(0)})}{P(Z > \frac{b-n_i\theta_0}{\sqrt{n_i\theta_0(1-\theta_0)}})} - 1 \right|$$

$$\leq \sup_{b \in \mathcal{D}_b} \left| \frac{P(\frac{\check{X}-n_i\theta_0}{\sqrt{n_i\theta_0(1-\theta_0)}} > 2\sqrt{\log m}) - P(Z > 2\sqrt{\log m})}{P(Z > \frac{b-n_i\theta_0}{\sqrt{n_i\theta_0(1-\theta_0)}})} \right|$$

$$+ \sup_{b \in \mathcal{D}_b} \left| \frac{P(\frac{b-n_i\theta_0}{\sqrt{n_i\theta_0(1-\theta_0)}} < \frac{\check{X}-n_i\theta_0}{\sqrt{n_i\theta_0(1-\theta_0)}} \leq 2\sqrt{\log m}) - P(2\sqrt{\log m} \geq Z > \frac{b-n_i\theta_0}{\sqrt{n_i\theta_0(1-\theta_0)}})}{P(Z > \frac{b-n_i\theta_0}{\sqrt{n_i\theta_0(1-\theta_0)}})} \right|$$

$\to 0$

Thus,

$$\sup_{b:|b-n_i\theta_0| \leq \sqrt{2n_i \log m \theta_0(1-\theta_0)}} \left| \frac{G_{0,i}^{(1)}(b + o(\sqrt{\frac{n}{\log m}}); \hat{c}^{(0)})}{G_{0,i}^{(1)}(b; \hat{c}^{(0)})} - 1 \right|$$

$$\leq \sup_{b:|b-n_i\theta_0| \leq \sqrt{2n_i \log m \theta_0(1-\theta_0)}} \left| \frac{P(Z > \frac{b-n_i\theta_0}{\sqrt{n_i\theta_0(1-\theta_0)}} + o(\sqrt{\frac{1}{\log m}}))}{P(Z > \frac{b-n_i\theta_0}{\sqrt{n_i\theta_0(1-\theta_0)}})} - 1 \right| + o(1)$$

$$= o(1)$$

When $b \leq n_i\theta_0 - \sqrt{2n_i \log m \theta_0(1-\theta_0)}$, we have

$$\sup_{b:b \leq n_i\theta_0 - \sqrt{2n_i \log m \theta_0(1-\theta_0)}} \left| \frac{G_{0,i}^{(1)}(b + o(\sqrt{\frac{n}{\log m}}); \hat{c}^{(0)})}{G_{0,i}^{(1)}(b; \hat{c}^{(0)})} - 1 \right|$$

$$\leq \frac{1 - G_{0,i}^{(1)}(n_i\theta_0 - \sqrt{2n_i \log m \theta_0(1-\theta_0)}; \hat{c}^{(0)})}{G_{0,i}^{(1)}(n_i\theta_0 - \sqrt{2n_i \log m \theta_0(1-\theta_0)}; \hat{c}^{(0)})}$$

$$= o(1)$$

Accordingly, we have

$$\sup_{b:b-n_i\theta_0\leq\sqrt{2n_i\log m\theta_0(1-\theta_0)}}\left|\frac{G_{0,i}^{(1)}(b+o(\sqrt{\frac{n}{\log m}});\hat{c}^{(0)})}{G_{0,i}^{(1)}(b;\hat{c}^{(0)})}-1\right|\to 0$$

$\square$

*Proof of Lemma 8.* For $i \in \mathcal{B}_{\text{nul}}^{(\ell)}$, when $\ell = 1$, we have

$$\frac{\mathsf{P}(\check{X}_i^{(1)}=\check{x}_i|\theta_i)}{\mathsf{P}(\check{X}_i^{(1)}=\check{x}_i|\theta_i)}=\left(\frac{\theta_i}{\theta_0}\right)^{x_i}\left(\frac{1-\theta_i}{1-\theta_0}\right)^{n_i-x_i}\leq\left[\left(\frac{\theta_i}{\theta_0}\right)^{\theta_0}\left(\frac{1-\theta_i}{1-\theta_0}\right)^{1-\theta_0}\right]^n$$

Given $\left(\frac{\theta_i}{\theta_0}\right)^{\theta_0}\left(\frac{1-\theta_i}{1-\theta_0}\right)^{1-\theta_0}$ is an increasing function of $\theta_i$ as long as $\theta_i \leq \theta_0$, we have

$$\frac{\mathsf{P}(\check{X}_i^{(1)}=\check{x}_i|\theta_i)}{\mathsf{P}(\check{X}_i^{(1)}=\check{x}_i|\theta_i)}\leq\left[\left(\frac{\theta_i}{\theta_0}\right)^{\theta_0}\left(\frac{1-\theta_i}{1-\theta_0}\right)^{1-\theta_0}\right]^n \leq 1$$

Thus, we have

$$\sup_{i \in \mathcal{B}_{nul}^{(1)}} |\tilde{G}_i^{(1)}(b^{(1)};c^{(0)}) - \tilde{G}_{0,i}^{(1)}(b^{(1)};c^{(0)})| \leq 0$$

When $\ell \geq 2$, we set $\check{X}_i^{(\ell)} = \check{X}_{i_1}^{(\ell)} + \check{X}_{i_2}^{(\ell)}$, with $\check{X}_{i_1}^{(\ell)} \sim \text{Binom}(n_{i_1},\theta_{i_1})$ and $\check{X}_{i_2}^{(\ell)} \sim \text{Binom}(n_{i_2},\theta_{i_2})$. By Condition C4, $\theta_{i_1} = \theta_{i_2} + o((n\log m)^{-1})$. Based on Taylor expansion,

$$\log[\theta_{i_2}^{\check{x}_2}(1-\theta_{i_2})^{n-\check{x}_2}] = \check{x}_2 \log\theta_{i_1} + (n_{i_1}-\check{x}_2)\log(1-\theta_{i_1}) + R_m$$

Here, $\sup_{i \in \mathcal{B}^{(\ell)}} R_m = o((\log m)^{-1})$, and accordingly,

$$\frac{\mathsf{P}(\check{X}_i^{(\ell)}=\check{x}|\theta_{i_1},\theta_{i_2})}{\mathsf{P}(\check{X}_i^{(\ell)}=\check{x}|\theta_{i_1}=\theta_{i_2}=\theta_0)}$$
$$=\frac{\sum_{\check{x}_1+\check{x}_2=x}\binom{n_{i_1}}{x_1}\binom{n_{i_2}}{x_2}\theta_{i_1}^{x_1}(1-\theta_{i_1})^{n_{i_1}-x_1}\theta_{i_2}^{x_2}(1-\theta_{i_2})^{n_{i_2}-x_2}}{\sum_{\check{x}_1+\check{x}_2=\check{x}}\binom{n_{i_1}}{x_1}\binom{n_{i_2}}{x_2}\theta_0^x(1-\theta_0)^{n_i-x}}$$
$$\leq \frac{\sum_{\check{x}_1+\check{x}_2=\check{x}}\binom{n_{i_1}}{x_1}\binom{n_{i_2}}{x_2}\theta_{i_1}^{\check{x}}(1-\theta_{i_1})^{n_i-\check{x}}}{\sum_{\check{x}_1+\check{x}_2=\check{x}}\binom{n_{i_1}}{x_1}\binom{n_{i_2}}{x_2}\theta_0^x(1-\theta_0)^{n_i-x}}\exp(\sup_{i \in \mathcal{B}^{(\ell)}} R_m)$$
$$\leq \exp(\sup_{i \in \mathcal{B}^{(\ell)}} R_m) = 1 + o(1)$$

When, $c^{(\ell-1)} \leq 1/2$, $b_{i_1}^{(\ell-1)} \geq n_{i_1}\theta_{i_1}$ and $b_{i_2}^{(\ell-1)} \geq n_{i_2}\theta_{i_2}$,

$$\widetilde{G}_i^{(\ell)}(b_i^{(\ell)}; c^{(\ell-1)}) = \frac{\mathsf{P}(\breve{X}_i \geq b^{(\ell)}, \breve{X}_{i_1}^{(\ell-1)} \leq b_{i_1}^{(\ell-1)}, \breve{X}_{i_2}^{(\ell-1)} \leq b_{i_2}^{(\ell-1)} \mid i \in \mathcal{B}_{\text{nul}}^{(\ell)})}{\{1 - \widetilde{G}_{i_1}^{(\ell-1)}(b_{i_1}^{(\ell-1)}; \hat{c}^{(0)})\}\{1 - \widetilde{G}_{i_2}^{(\ell-1)}(b_{i_2}^{(\ell-1)}; \hat{c}^{(0)})\}}$$

$$\leq \frac{\mathsf{P}(\breve{X}_i \geq b_i^{(\ell)}, \breve{X}_{i_1}^{(\ell-1)} \leq b_{i_1}^{(\ell-1)}, \breve{X}_{i_2}^{(\ell-1)} \leq b_{i_2}^{(\ell-1)} \mid \theta_j = \theta_0, \forall j \in S_i^{(\ell)})}{\{1 - G_0^{(\ell-1)}(b_{i_1}^{(\ell-1)}; \hat{c}^{(0)})\}\{1 - G_0^{(\ell-1)}(b_{i_2}^{(\ell-1)}; \hat{c}^{(0)})\}} = G_{0,i}^{(\ell)}(b_i^{(\ell)}; c^{(\ell-1)}).$$

Therefore, (S6) holds.

Following the similar argument, we can prove (S7)–(S9). □

*Proof of Lemma 9.* Given (b) follows immediately after (a), we will focus on the proof of (a).

Let $Z_1, Z_2, ..., Z_{2^{\ell-1}} \overset{iid}{\sim} N(0,1)$, $\tau^{(\ell)} = \frac{b^{(\ell)} - n^{(\ell)}\theta_0}{\sqrt{\{n^{(\ell)}\theta_0(1-\theta_0)\}}}$, $\widetilde{X}_{k_1} = \frac{\sum_{i=1}^{2^{k-1}} \breve{X}_i - n^{(k)}\theta_0}{\sqrt{n^{(k)}\theta_0(1-\theta_0)}}$, $\widetilde{X}_{k_2} = \frac{\sum_{i=2^{k-1}+1}^{2^{\ell-1}} \breve{X}_i - (n^{(\ell)}-n^{(k)})\theta_0}{\sqrt{(n^{(\ell)}-n^{(k)})\theta_0(1-\theta_0)}}$. When $\tau_i^{(\ell)} > \beta_0/\sqrt{2^{\ell-k}}$,

$$P(\sum_{i=1}^{2^{\ell-1}} \breve{X}_i > b^{(\ell)}, \sum_{i=1}^{2^{k-1}} \breve{X}_i > b_k)$$

$$= P(\frac{1}{\sqrt{2^{\ell-k}}}\widetilde{X}_{k_1} + \sqrt{1 - \frac{1}{2^{\ell-k}}}\widetilde{X}_{k_2} > \tau^{(\ell)}, \widetilde{X}_{k_1} > \beta_0)$$

$$\leq \mathsf{P}(\frac{1}{\sqrt{2^{\ell-k}}}\widetilde{X}_{k_1} + \sqrt{1 - \frac{1}{2^{\ell-k}}}\widetilde{X}_{k_2} > \tau^{(\ell)}, \beta_0 < \widetilde{X}_{k_1} \leq \sqrt{2^{\ell-k}}\gamma, -\gamma < \widetilde{X}_{k_2} \leq \left[1 - \frac{1}{2^{\ell-k}}\right]^{-1/2}\gamma)$$

$$+ \mathsf{P}(\widetilde{X}_{k_1} > \sqrt{2^{\ell-k}}\gamma) + \mathsf{P}(\widetilde{X}_{k_2} > \left[1 - \frac{1}{2^{\ell-k}}\right]^{-1/2}\gamma) \quad (S43)$$

By applying the Lemma 4 on the first term in (S43),

$$\mathsf{P}(\frac{1}{\sqrt{2^{\ell-k}}}\widetilde{X}_{k_1} + \sqrt{1 - \frac{1}{2^{\ell-k}}}\widetilde{X}_{k_2} > \tau^{(\ell)}, \beta_0 < \widetilde{X}_{k_1} \leq \sqrt{2^{\ell-k}}\gamma, -\gamma < \widetilde{X}_{k_2} \leq \left[1 - \frac{1}{2^{\ell-k}}\right]^{-1/2}\gamma)$$

$$\leq 2\mathsf{P}(\frac{1}{\sqrt{2^{\ell-k}}}Z_1 + \sqrt{1 - \frac{1}{2^{\ell-k}}}Z_2 > \tau^{(\ell)}, \beta_0 < Z_1 \leq \sqrt{2^{\ell-k}}\gamma, -\gamma < Z_2 \leq \left[1 - \frac{1}{2^{\ell-k}}\right]^{-1/2}\gamma)$$

$$\leq 2\mathsf{P}(\frac{1}{\sqrt{2^{\ell-k}}}Z_1 + \sqrt{1 - \frac{1}{2^{\ell-k}}}Z_2 > \tau^{(\ell)}, Z_1 > \beta_0)$$

Thus, based on the Lemma 5, when $m$ and $n$ are large sufficiently, we have

$$\frac{P(\sum_{i=1}^{2^{\ell-1}} \breve{X}_i > b^{(\ell)}, \sum_{i=1}^{2^{k-1}} \breve{X}_i > b_k)}{\mathsf{P}(\sum_{i=1}^{2^{\ell-1}} \breve{X}_i > b^{(\ell)})}$$

$$\leq \frac{4\mathsf{P}(\frac{1}{\sqrt{2^{\ell-k}}}Z_1 + \sqrt{1 - \frac{1}{2^{\ell-k}}}Z_2 > \tau^{(\ell)}, Z_1 > \beta_0)}{\mathsf{P}(\frac{1}{\sqrt{2^{\ell-k}}}Z_1 + \sqrt{1 - \frac{1}{2^{\ell-k}}}Z_2 > \tau^{(\ell)})} + o(1) \quad (S44)$$

In addition, when $\tau_i^{(\ell)} \leq \beta_0/\sqrt{2^{\ell-k}}$, we have

$$\max_{k=1,\ldots,\ell-1} \frac{P(\sum_{i=1}^{2^{\ell-1}} \breve{X}_i > b^{(\ell)}, \sum_{i=1}^{2^{k-1}} \breve{X}_i > b_k)}{\mathsf{P}(\sum_{i=1}^{2^{\ell-1}} \breve{X}_i > b^{(\ell)})}$$
$$\leq \max_{k=1,\ldots,\ell-1} \frac{2\mathsf{P}(Z_1 > \beta_0)}{\mathsf{P}(\frac{1}{\sqrt{2^{\ell-k}}}Z_1 + \sqrt{1-\frac{1}{2^{\ell-k}}}Z_2 > \beta_0/\sqrt{2^{\ell-k}})} \to 0$$

Thus, the prove for (S10) is finished when

$$\lim_{m\to\infty} \max_{k=1,\ldots,\ell-1} \sup_{\tau^{(\ell)}\in(\beta_0/\sqrt{2^{\ell-k}},\gamma]} \frac{\mathsf{P}(\frac{1}{\sqrt{2^{\ell-k}}}Z_1 + \sqrt{1-\frac{1}{2^{\ell-k}}}Z_2 > \tau^{(\ell)}, Z_1 > \beta_0)}{\mathsf{P}(\frac{1}{\sqrt{2^{\ell-k}}}Z_1 + \sqrt{1-\frac{1}{2^{\ell-k}}}Z_2 > \tau^{(\ell)})} = 0$$

For each $m$, define $\mathcal{D}_m = \{\tau^{(\ell)} \in (\beta_0/\sqrt{2^{\ell-k}},\gamma) : \frac{d}{d\tau^{(\ell)}} \frac{\mathsf{P}(\frac{1}{\sqrt{2^{\ell-k}}}Z_1+\sqrt{1-\frac{1}{2^{\ell-k}}}Z_2>\tau^{(\ell)},Z_1>\beta_0)}{\mathsf{P}(\frac{1}{\sqrt{2^{\ell-k}}}Z_1+\sqrt{1-\frac{1}{2^{\ell-k}}}Z_2>\tau^{(\ell)})} = 0\}$, then

$$\sup_{\tau^{(\ell)}\in(\beta_0/\sqrt{2^{\ell-k}},\gamma]} \frac{\mathsf{P}(\frac{1}{\sqrt{2^{\ell-k}}}Z_1 + \sqrt{1-\frac{1}{2^{\ell-k}}}Z_2 > \tau^{(\ell)}, Z_1 > \beta_0)}{\mathsf{P}(\frac{1}{\sqrt{2^{\ell-k}}}Z_1 + \sqrt{1-\frac{1}{2^{\ell-k}}}Z_2 > \tau^{(\ell)})}$$
$$\leq \max\left\{ \sup_{\tau^{(\ell)}=\beta_0/\sqrt{2^{\ell-k}} \text{ or } \gamma} \frac{\mathsf{P}(\frac{1}{\sqrt{2^{\ell-k}}}Z_1 + \sqrt{1-\frac{1}{2^{\ell-k}}}Z_2 > \tau^{(\ell)}, Z_1 > \beta_0)}{\mathsf{P}(\frac{1}{\sqrt{2^{\ell-k}}}Z_1 + \sqrt{1-\frac{1}{2^{\ell-k}}}Z_2 > \tau^{(\ell)})}, \right.$$
$$\left. \sup_{\tau^{(\ell)}\in\mathcal{D}_m} \frac{\mathsf{P}(\frac{1}{\sqrt{2^{\ell-k}}}Z_1 + \sqrt{1-\frac{1}{2^{\ell-k}}}Z_2 > \tau^{(\ell)}, Z_1 > \beta_0)}{\mathsf{P}(\frac{1}{\sqrt{2^{\ell-k}}}Z_1 + \sqrt{1-\frac{1}{2^{\ell-k}}}Z_2 > \tau^{(\ell)})} \right\}$$

(i). When $\tau^{(\ell)} = \beta_0/\sqrt{2^{\ell-k}}$ or $\gamma$, by L'Hopital's rule, we have

$$\lim_{m\to\infty} \max_{k=1,\ldots,\ell-1} \frac{\mathsf{P}(\frac{1}{\sqrt{2^{\ell-k}}}Z_1 + \sqrt{1-\frac{1}{2^{\ell-k}}}Z_2 > \tau^{(\ell)}, Z_1 > \beta_0)}{\mathsf{P}(\frac{1}{\sqrt{2^{\ell-k}}}Z_1 + \sqrt{1-\frac{1}{2^{\ell-k}}}Z_2 > \tau^{(\ell)})} = 0$$

(ii). When $\tau^{(\ell)} \in \mathcal{D}_m$, given

$$0 = \frac{d}{d\tau^{(\ell)}} \frac{\mathsf{P}(\frac{1}{\sqrt{2^{\ell-k}}}Z_1 + \sqrt{1-\frac{1}{2^{\ell-k}}}Z_2 > \tau^{(\ell)}, Z_1 > \beta_0)}{\mathsf{P}(\frac{1}{\sqrt{2^{\ell-k}}}Z_1 + \sqrt{1-\frac{1}{2^{\ell-k}}}Z_2 > \tau^{(\ell)})}$$
$$= \frac{\mathsf{P}(\frac{1}{\sqrt{2^{\ell-k}}}Z_1 + \sqrt{1-\frac{1}{2^{\ell-k}}}Z_2 > \tau^{(\ell)})\frac{d}{d\tau^{(\ell)}}\mathsf{P}(\frac{1}{\sqrt{2^{\ell-k}}}Z_1 + \sqrt{1-\frac{1}{2^{\ell-k}}}Z_2 > \tau^{(\ell)}, Z_1 > \beta_0)}{\mathsf{P}(\frac{1}{\sqrt{2^{\ell-k}}}Z_1 + \sqrt{1-\frac{1}{2^{\ell-k}}}Z_2 > \tau^{(\ell)})^2}$$
$$- \frac{\mathsf{P}(\frac{1}{\sqrt{2^{\ell-k}}}Z_1 + \sqrt{1-\frac{1}{2^{\ell-k}}}Z_2 > \tau^{(\ell)}, Z_1 > \beta_0)\frac{d}{d\tau^{(\ell)}}\mathsf{P}(\frac{1}{\sqrt{2^{\ell-k}}}Z_1 + \sqrt{1-\frac{1}{2^{\ell-k}}}Z_2 > \tau^{(\ell)})}{\mathsf{P}(\frac{1}{\sqrt{2^{\ell-k}}}Z_1 + \sqrt{1-\frac{1}{2^{\ell-k}}}Z_2 > \tau^{(\ell)})^2}$$

We have

$$\frac{\mathsf{P}(\frac{1}{\sqrt{2^{\ell-k}}}Z_1 + \sqrt{1 - \frac{1}{2^{\ell-k}}}Z_2 > \tau^{(\ell)}, Z_1 > \beta_0)}{\mathsf{P}(\frac{1}{\sqrt{2^{\ell-k}}}Z_1 + \sqrt{1 - \frac{1}{2^{\ell-k}}}Z_2 > \tau^{(\ell)})}$$
$$= \frac{\frac{d}{d\tau^{(\ell)}}\mathsf{P}(\frac{1}{\sqrt{2^{\ell-k}}}Z_1 + \sqrt{1 - \frac{1}{2^{\ell-k}}}Z_2 > \tau^{(\ell)}, Z_1 > \beta_0)}{\frac{d}{d\tau^{(\ell)}}\mathsf{P}(\frac{1}{\sqrt{2^{\ell-k}}}Z_1 + \sqrt{1 - \frac{1}{2^{\ell-k}}}Z_2 > \tau^{(\ell)})}$$

Therefore,

$$\max_{k=1,\ldots,\ell-1} \sup_{\tau^{(\ell)} \in \mathcal{D}_m} \frac{\mathsf{P}(\frac{1}{\sqrt{2^{\ell-k}}}Z_1 + \sqrt{1 - \frac{1}{2^{\ell-k}}}Z_2 > \tau^{(\ell)}, Z_1 > \beta_0)}{\mathsf{P}(\frac{1}{\sqrt{2^{\ell-k}}}Z_1 + \sqrt{1 - \frac{1}{2^{\ell-k}}}Z_2 > \tau^{(\ell)})}$$
$$= \max_{k=1,\ldots,\ell-1} \sup_{\tau^{(\ell)} \in \mathcal{D}_m} \frac{\frac{d}{d\tau^{(\ell)}}\mathsf{P}(\frac{1}{\sqrt{2^{\ell-k}}}Z_1 + \sqrt{1 - \frac{1}{2^{\ell-k}}}Z_2 > \tau^{(\ell)}, Z_1 > \beta_0)}{\frac{d}{d\tau^{(\ell)}}\mathsf{P}(\frac{1}{\sqrt{2^{\ell-k}}}Z_1 + \sqrt{1 - \frac{1}{2^{\ell-k}}}Z_2 > \tau^{(\ell)})}$$
$$\to 0$$

By combining (i) and (ii),

$$\lim_{m \to \infty} \max_{k=1,\ldots,\ell-1} \sup_{\tau^{(\ell)} \in (\beta_0/\sqrt{2^{\ell-k}}, \gamma]} \frac{\mathsf{P}(\frac{1}{\sqrt{2^{\ell-k}}}Z_1 + \sqrt{1 - \frac{1}{2^{\ell-k}}}Z_2 > \tau^{(\ell)}, Z_1 > \beta_0)}{\mathsf{P}(\frac{1}{\sqrt{2^{\ell-k}}}Z_1 + \sqrt{1 - \frac{1}{2^{\ell-k}}}Z_2 > \tau^{(\ell)})} = 0$$

□

*Proof of Lemma 10.* If $\hat{c}^{(\ell)} \leq a_N^{(\ell)}$, all statements hold immediately. Now we prove the lemma when $\hat{c}^{(\ell)} > a_N^{(\ell)}$.

**Step 1: Prove that on Layer 1, (a) holds**

Let $b_{-1} < b_0 < \ldots < b_{\lceil \gamma(\log m \log \log m)^{1/2} \rceil}$ be the values defined in the proof of Theorem 1. We can get the corresponding p-values sequence $q_{-1}^{(1)} > \ldots > q_{\lceil \gamma(\log m \log \log m)^{1/2} \rceil}^{(1)}$, under Binom$(n, \theta_0)$. Let value $q' = \frac{C^{(1)} \log m}{m}$. Based on (S15), $P(\hat{c}^{(1)} \geq q') \to 1$. We define the working sequence on layer 1 as $Q_{sub}^{(1)} = \{q_{-1}^{(1)}, \ldots, q_t^{(1)}, q'\}$, where $t \in \{0, \ldots, \lceil \gamma(\log m \log \log m)^{1/2} \rceil - 1\}$ is the index s.t. $q_t^{(1)} \geq q'$ and $q_{t+1}^{(1)} \leq q'$.

By Markov Inequality,

$$\mathsf{P}\left[\max_{q\in Q_{sub}^{(1)}}\left|\frac{\sum_{i\in\mathcal{B}_{\text{nul}}^{(1)}}I(P_{0,i}^{(1)}<q)-\sum_{i\in\mathcal{B}_{\text{nul}}^{(1)}}\mathsf{P}(P_{0,i}^{(1)}<q)}{\left\{\sum_{i\in\mathcal{B}_{\text{nul}}^{(1)}}\mathsf{P}(P_{0,i}^{(1)}<q)\right\}^{1/2}}\right|>\left\{\log m^{(1)}\right\}^{1.6/4}\right]$$

$$\leq C(\log m)^{3/2}\frac{E\left[(\sum_{i\in\mathcal{B}_{\text{nul}}^{(1)}}I(P_{0,i}^{(1)}<q)-\sum_{i\in\mathcal{B}_{\text{nul}}^{(1)}}\mathsf{P}(P_{0,i}^{(1)}<q))^{4}\right]}{\left[\sum_{i\in\mathcal{B}_{\text{nul}}^{(1)}}\mathsf{P}(P_{0,i}^{(1)}<q)\right]^{2}(\log m)^{1.6}}$$

$$\leq C(\log m)^{3/2}\frac{\left[m_0 q(1-q)\left(1-6q(1-q)\right)+m_0^2 q^2(1-q)^2\right]\left(1+C\frac{16n^2}{N}\right)}{m_0^2 q^2(1+o(1))(\log m)^{1.6}} \quad \text{(Lemma 1)}$$

$$\leq C(\log m)^{-0.1} \tag{S45}$$

Thus, by Lemma 3 and (S15),

$$\mathsf{P}\left[\max_{q\in Q_{sub}^{(1)}}\left|\frac{\sum_{i\in\mathcal{B}_{\text{nul}}^{(1)}}I(P_{0,i}^{(1)}<q)-\sum_{i\in\mathcal{B}_{\text{nul}}^{(1)}}\mathsf{P}(P_{0,i}^{(1)}<q)}{\left\{\sum_{i\in\mathcal{B}_{\text{nul}}^{(1)}}\mathsf{P}(P_{0,i}^{(1)}<q)\right\}}\right|>\frac{1}{C^{(1)}\left\{\log m^{(1)}\right\}^{0.1}}\right]$$

$$\to 0 \tag{S46}$$

Given

$$\sup_{j=-1,\ldots,t}\left|\frac{q_j}{q_{j+1}}-1\right|\leq\sup_{j=-1,\ldots,t}\left|\frac{G_{0,i}^{(1)}(b_j;\hat{c}^{(0)})}{G_{0,i}^{(1)}(b_{j+1};\hat{c}^{(0)})}-1\right|+o(1)=o(1)$$

Combined with (S27), (S45) and (S46), there exists $\delta_2(m)=o(1)$, s.t.

$$\mathsf{P}\left[\sup_{q\in[q',1/2]}\left|\frac{\sum_{i\in\mathcal{B}_{\text{nul}}^{(1)}}I(P_{0,i}^{(1)}<q)-\sum_{i\in\mathcal{B}_{\text{nul}}^{(1)}}P(P_{0,i}^{(1)}<q)}{\left\{\sum_{i\in\mathcal{B}_{\text{nul}}^{(1)}}\mathsf{P}(P_{0,i}^{(1)}<q)\right\}}\right|>\delta_2(m)\right]\to 0$$

and accordingly,

$$\mathsf{P}\left[\left|\frac{\sum_{i\in\mathcal{B}_{\text{nul}}^{(1)}}I(P_{0,i}^{(1)}<\hat{c}^{(1)})-\sum_{i\in\mathcal{B}_{\text{nul}}^{(1)}}\mathsf{P}(P_{0,i}^{(1)}<\hat{c}^{(1)})}{\left\{\sum_{i\in\mathcal{B}_{\text{nul}}^{(1)}}\mathsf{P}(P_{0,i}^{(1)}<\hat{c}^{(1)})\right\}}\right|>\delta_2(m)\right]\to 0 \tag{S47}$$

**Step 2: Prove statement (b)**
From (7), we have

$$m^{(\ell)}\hat{c}^{(\ell)}\leq\alpha\max\left\{\sum_{i=1}^{m^{(\ell)}}I(P_{0,i}^{(\ell)}<\hat{c}^{(\ell)}),1\right\} \tag{S48}$$

We also know that

$$\sum_{i=1}^{m^{(\ell)}}I(P_{0,i}^{(\ell)}<\hat{c}^{(\ell)})=\sum_{i\in\mathcal{B}_{\text{nul}}^{(\ell)}}I(P_{0,i}^{(\ell)}<\hat{c}^{(\ell)})+\sum_{i\in\mathcal{B}_{\text{alt}}^{(\ell)}}I(P_{0,i}^{(\ell)}<\hat{c}^{(\ell)})$$

$$\leq\sum_{i\in\mathcal{B}_{\text{nul}}^{(\ell)}}I(P_{0,i}^{(\ell)}<\hat{c}^{(\ell)})+(m^{(\ell)})^{r_1} \tag{S49}$$

In view of Lemma 8, on $\cap_{h=1}^{\ell} \mathcal{X}^{(h)}$,

$$\sum_{i \in \mathcal{B}_{\text{nul}}^{(\ell)}} I(P_{0,i}^{(\ell)} < \hat{c}^{(\ell)})$$
$$\leq \sum_{i \in \mathcal{B}_{\text{nul}}^{(\ell)}} \mathsf{P}(P_{0,i}^{(\ell)} < \hat{c}^{(\ell)}) + \left\{ \sum_{i \in \mathcal{B}_{\text{nul}}^{(\ell)}} \mathsf{P}(P_{0,i}^{(\ell)} < \hat{c}^{(\ell)}) \right\} \delta_2(m)$$
$$\leq m_0^{(\ell)} \hat{c}^{(\ell)} (1 + \delta_2(m)) \tag{S50}$$

By combining (S48), (S49) and (S50),

$$(1 - \alpha(1 + o(1))) m_0^{(\ell)} \hat{c}^{(\ell)} \leq \alpha (m^{(\ell)})^{r_1}$$

It is easy to see that

$$\hat{c}^{(\ell)} \leq C (m^{(\ell)})^{r_1 - 1}, \tag{S51}$$

**Step 3: Prove statement (c) on $\cap_{h=1}^{\ell-1} \mathcal{X}^{(h)}$**

We start to prove (S17) by induction. When $\ell = 2$, based on (S16), $G_{0,i}^{(1)}(\hat{b}_i^{(1)}; \hat{c}^{(0)}) \leq C(m^{(1)})^{r_1-1}$. Then by Lemma 5, we know that $\hat{\tau}_i^{(1)} \geq \beta\{1 + o(1)\}$.

Now suppose $\hat{\tau}_i^{(k)} \geq \beta\{1 + o(1)\}$ for $k = 1, \ldots, \ell - 2$. On layer $\ell - 1$,

$$G_{0,i}^{(\ell-1)}(\hat{b}_i^{(\ell-1)}; \hat{c}^{(0)})$$
$$\leq G_{0,i}^{(\ell-1)}(\hat{b}_i^{(\ell-1)}; \hat{c}^{(\ell-2)}) \{1 - \widetilde{G}_{0,i}^{(\ell-2)}(\hat{b}_i^{(\ell-2)}; \hat{c}^{(0)})\}^2 + 3 G_{0,i}^{(\ell-2)}(\hat{b}_i^{(\ell-2)}; \hat{c}^{(0)}) \tag{S52}$$

by the right-handed side of (S52) and (S51), $G_{0,i}^{(\ell-1)}(\hat{b}_i^{(\ell-1)}; \hat{c}^{(0)}) \leq C(m^{(1)})^{r_1-1} + 3C(m^{(1)})^{r_1-1} \leq C(m^{(1)})^{r_1-1}$. Then by Lemma 5, $\hat{\tau}_i^{(\ell-1)} \geq \beta\{1 + o(1)\}$.

**Step 4: Prove that (a) holds on layer $\ell$ when it holds on layer $1, ..., \ell-1$**

Let $b_{-1} < b_0 < ... < b_{\lceil \gamma(\log m \log \log m)^{1/2} \rceil}$ be the values defined in the proof of Theorem 1. In view of Lemma 9 (b),

$$\sup_{t=0,\ldots,\lceil \gamma(\log m \log \log m)^{1/2}\rceil} \left| \frac{G_{0,i}^{(\ell)}(b_t; \hat{c}^{(\ell-1)})}{G_{0,i}^{(\ell)}(b_{t-1}; \hat{c}^{(\ell-1)})} - 1 \right| = o(1)$$

By following the similar arguments in Step 1, we have $P(\mathcal{X}^{(\ell)}) \to 1$. Thus,

$$P(\cap_{h=1}^{\ell} \mathcal{X}^{(h)}) \geq 1 - o(1)$$

□

*Proof of Lemma 11.* Based on (7), we know that

$$m_0^{(1)} G_0^{(1)}(\hat{b}_i^{(1)} - 1; \hat{c}^{(0)}) > \alpha \max \left\{ \sum_{i=1}^{m^{(\ell)}} I(X_i^{(1)} > \hat{b}_i^{(1)} - 1), 1 \right\}$$
$$\geq \alpha \max \left\{ \sum_{i=1}^{m^{(\ell)}} I(X_i^{(1)} > \hat{b}_i^{(1)}), 1 \right\} \tag{S53}$$

By Lemma 8,

$$\begin{aligned}
&G_{0,i}^{(\ell)}(\hat{b}_i^{(\ell)} - 1; \hat{c}^{(\ell-1)}) \\
&\leq G_{0,i}^{(\ell)}(\hat{b}_i^{(\ell)}; \hat{c}^{(\ell-1)}) + g_{0,i}^{(\ell)}(\hat{b}_i^{(\ell)}; \hat{c}^{(\ell-1)}) \\
&\leq G_{0,i}^{(\ell)}(\hat{b}_i^{(\ell)}; \hat{c}^{(\ell-1)}) + \frac{g_{0,i}^{(\ell)}(\hat{b}_i^{(\ell)}; \hat{c}^{(0)})}{\{1 - G_{0,i}^{(\ell-1)}(\hat{b}_{i_1}^{(\ell-1)}; \hat{c}^{(0)})\}\{1 - G_{0,i}^{(\ell-1)}(\hat{b}_{i_2}^{(\ell-1)}; \hat{c}^{(0)})\}}
\end{aligned}$$

By Lemma 4 and Lemma 5, $g_{0,i}^{(\ell)}(\hat{b}_i^{(\ell)}; \hat{c}^{(0)}) \asymp G_{0,i}^{(\ell)}(\hat{b}_i^{(\ell)}; \hat{c}^{(0)})\hat{\tau}^{(\ell)}n^{-1/2}$. By (7), we know that $\tau_i^{(\ell)} n^{-1/2} \leq C(\log m^{(1)})^{1/2} n^{-1/2} = o(1)$. Combined with (S13), we know that on $\mathcal{X}$,

$$G_{0,i}^{(\ell)}(\hat{b}_i^{(\ell)} - 1; \hat{c}^{(\ell-1)}) \leq G_{0,i}^{(\ell)}(\hat{b}_i^{(\ell)}; \hat{c}^{(\ell-1)}) \left\{1 + C(\log m^{(1)})^{1/2} n^{-1/2}\right\} \tag{S54}$$

Thus on $\mathcal{X}$,

$$m_0^{(\ell)} G_{0,i}^{(\ell)}(\hat{b}_i^{(\ell)}; \hat{c}^{(\ell-1)}) \leq \alpha \max\left\{\sum_{i=1}^{m^{(\ell)}} I(X_i^{(\ell)} > \hat{b}_i^{(\ell)}), 1\right\}$$
$$\leq m_0^{(\ell)} G_{0,i}^{(\ell)}(\hat{b}_i^{(\ell)}; \hat{c}^{(\ell-1)}) \left\{1 + C(\log m^{(1)})^{1/2} n^{-1/2}\right\}.$$

And therefore,

$$\frac{m_0^{(\ell)} G_{0,i}^{(\ell)}(\hat{b}_i^{(\ell)}; \hat{c}^{(\ell-1)})}{\max\left\{\sum_{i=1}^{m^{(\ell)}} I(X_i^{(\ell)} > \hat{b}_i^{(\ell)}), 1\right\}} = \alpha \left\{1 + C(\log m^{(1)})^{1/2} n^{-1/2}\right\}.$$

$\square$

**S5. Univariate density plots of functional markers by activation status (based on TEAM results) and functional group.**

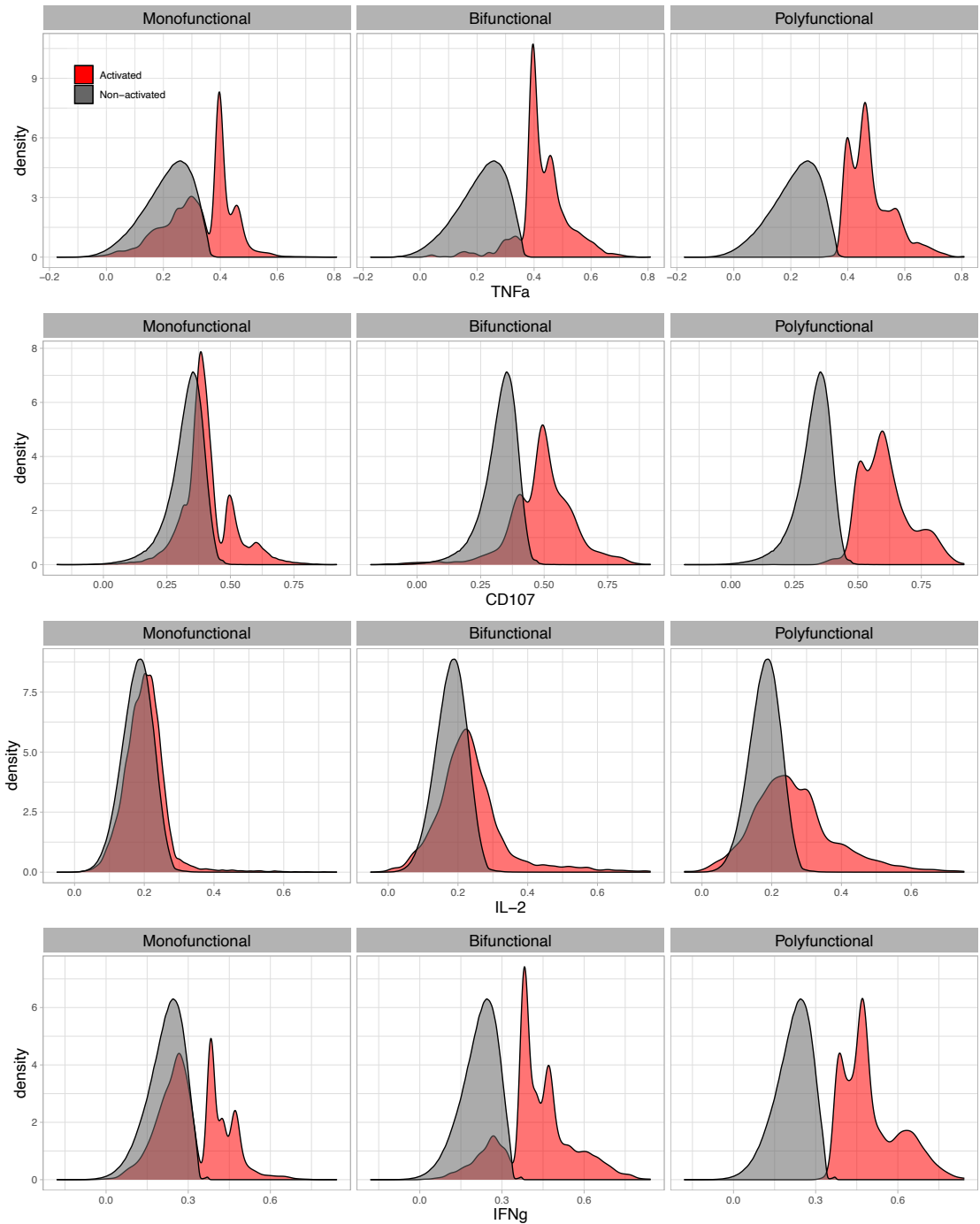

Supplemental Material, Figure S1: Univariate density plots of functional markers by activation status (based on TEAM results) and functional group. A general trend that can be observed is that as cells move from monofunctional to bifunctional to polyfunctional, the average expression level of each functional marker increases. As the fluorescence intensity of each marker is transformed to an approximate log-scale, small changes along the x-axis represent large changes in the number of marker molecules on a cell. The emergence of distinct modes in the density plots also suggests that there are discrete subsets of cell activation states.